\documentclass{article}

\usepackage{arxiv}

\usepackage[utf8]{inputenc} 
\usepackage[T1]{fontenc}    
\usepackage{hyperref}       
\usepackage{url}            
\usepackage{booktabs}       
\usepackage{amsfonts}       
\usepackage{nicefrac}       
\usepackage{microtype}      
\usepackage{lipsum}		
\usepackage{graphicx}
\usepackage{natbib}
\usepackage{doi}
\usepackage{amsmath}
\usepackage{theorem}
\usepackage{dsfont}
\usepackage{float}

\usepackage{tikz}
\usetikzlibrary{positioning}
\usepackage{adjustbox}  

\def\R{\mathbb{R}}
\def\calN{\mathcal{N}}

\newtheorem{Theo}{Theorem}
\newtheorem{Ass}{Assumption}
\newtheorem{Rem}{Remark}
\newtheorem{example}{Example}
\newtheorem{Lemma}{Lemma}

\title{Modelling failure risks in load sharing systems with
heterogeneous components}

\author{Tim Pesch, Adriano Polpo, Edward Cripps\thanks{T. Pesch, A. Polpo and E. Cripps are with the ARC Centre for Transforming Maintenance through Data Science and the Department of Mathematics and Statistics at the University of Western Australia, Perth, Australia.}~ and Erhard Cramer\thanks{E. Cramer is with the Institute of Statistics at
RWTH Aachen University, Aachen, Germany.}
}

\renewcommand{\headeright}{}
\renewcommand{\undertitle}{}
\renewcommand{\shorttitle}{Load sharing systems with heterogeneous components}

\hypersetup{
pdftitle={MODELLING FAILURE RISKS IN LOAD SHARING SYSTEMS WITH
HETEROGENEOUS COMPONENTS},
pdfsubject={q-bio.NC, q-bio.QM},
pdfauthor={Tim~Pesch, Adriano~Polpo, Edward~Cripps, Erhard~Cramer}
}

\begin{document}
\maketitle

\begin{abstract}
	A load sharing system has several components and the failure of one component can affect the lifetime of the surviving components. Since component failure does not equate to system failure for different system designs, the analysis of the dependency structure between components becomes a meaningful exercise. The Extended Sequential Order Statistics model allows us to model a dependence structure between heterogeneous components in load sharing systems. However, the results may suggest that the risk of failure decreases as components fail sequentially, which can be counterintuitive, especially when data is scarce. We propose to address this issue by imposing an order restriction on the model parameters that represent increasing failure risks. This assumption corresponds more realistically to the physical properties of the system in many applications. We discuss the advantages of the newly proposed estimates and describe situations where they should be used with caution.
\end{abstract}


\section{Introduction}

In 1995, \cite{Kamps1995} proposed the sequential order statistics' (SOS) model as an extension to order statistics (OS). In contrast to ordinary OS, SOS allow us to model a dependence structure between components in load sharing systems. More precisely, SOS account for the impact of component failures on the remaining, still operating components. For example, a series of wash tanks that filter an influx of contaminated product is an example of a system with a cause-and-effect relationship between components. The failure of one wash tank will usually increase the load on the remaining tanks. Typically, this increase is assumed to result in a reduced lifetime of the surviving components. For more recent developments which take a different approach to modelling load sharing systems, we refer the reader to \cite{Zhao2018}, \cite{Sutar2014}, and \cite{Mueller2022}. For references on SOS as load sharing model, see \cite{Balakrishnan2011,Bedbur2019, Mies2019}.

The SOS model originally assumes all component lifetimes to be identically distributed, a simplification that limits the model's applicability in real-life scenarios. Many applications would benefit from greater flexibility, for instance, when components are either of different type, i.e. they are different from one another regarding their technical specifications, or when components fulfil various tasks within the system. An example of the latter is a series system in which components experience different workloads based on their position in the series. Components may have different lifetime expectations while possibly being of the same type. This limitation is addressed in \cite{Baratnia2017}. The authors relax the assumption of identically distributed lifetimes and introduce heterogeneous components instead. The resulting model, the Extended Sequential Order Statistics (ESOS) model, is better suited in situations where components differ in either type or function within the system.

Both models, SOS and ESOS have led to interesting estimation and hypothesis testing methods. Some recent results based on ESOS include the derivation of maximum likelihood estimates (MLEs) of the parameters of the underlying exponential lifetime distributions. These estimates have desirable properties, including unbiasedness and consistency (see \cite{Pesch2023}). Another desirable property of the estimates is to support the idea of an increasing risk of failure as components break successively. Typically, this belief is justified since the failure of one component increases the stress on the remaining components. We will demonstrate that the MLEs proposed in \cite{Pesch2023} may suggest reduced risks of failure, especially if sample sizes are small, which contradicts the physical property of the system. To address this issue, we propose a new set of estimates to overcome counterintuitive results, which we call order restricted (OR) estimates. These yield results which agree with our intuition of increasing failure risks. This assumption is feasible for most technical load sharing systems. Upon the failure of one component, the additional load needs to be distributed among the surviving components. We demonstrate the usefulness of the OR estimates with the help of examples of such load sharing systems for which we assume a k-out-of-n:F design. In this design, a system is considered broken if and only if at least $k$ of its $n$ components have failed.

The article is structured as follows: background information on the models of SOS and ESOS is provided in Section \ref{Sec-0}. In Section \ref{Sec-1}, we introduce the Conditional Proportional Hazard Rate (CPHR) model for ESOS and present an expression for the MLEs of the model parameters. We also derive expressions for the MLEs of the model parameters when the exact failure history is assumed irrelevant but only the current state of the system matters (Section \ref{Sec-2}). In Section \ref{Sec-3}, the need for better-suited estimates is motivated via a data example. As a solution to the problem, we state the proposed OR MLEs. The properties of the newly proposed set of estimates are thoroughly investigated with an extensive simulation studies in Section \ref{Sec-4}. In Section \ref{Sec-5}, we establish explicit expressions for the OR MLEs in case of some parametric families of life time distributions.

\section{Model and Notation}
\label{Sec-0}

The lifetimes of individual components of multi-component systems are historically often modelled as independent and identically distributed. The model of SOS relaxes this assumption by permitting a dependence structure between components, thereby increasing the model's flexibility. Upon failure of one component, the stress on the remaining components increases, and their respective lifetime expectation consequently reduces. The increasingly ordered failure times are denoted by $X_*^{(1)},\dots,X_*^{(s)}$ and are called SOS (\cite{Kamps1995}). For an extensive insight into the mostly technical construction of SOS and their relation to other models of ordered random variables, see \cite{Cramer1999d, Cramer2001a,Cramer2003}. For additional reading on failures inducing an increased failure rate on the surviving components, we refer to \cite{Scheuer1988}. New and meaningful results quickly emerged from the model. In \cite{Cramer1996}, the MLEs for the underpinning model parameters are derived based on the joint density of the first $s\leq n$ SOS $X_*^{(1)} \leq \dots \leq X_*^{(s)}$, i.e.,
\begin{align}
\label{eq:SOS}
f_{X_*^{(1)},\dots,X_*^{(s)}}(x_1,\dots,x_s) &= \frac{n!}{(n-s)!}\big(1-F_s(x_{s})\big)^{n-s} f_s(x_{s}) \prod_{i=1}^{s-1}\bigg(\frac{1-F_i(x_{i})}{1-F_{i+1}(x_{i})} \bigg)^{n-i} f_i(x_{i}),
\end{align}
where $n$ is the system size and $F_i$ and $f_i, i\in \{1,\dots,n\}=\mathcal{N}$, say, respectively. The construction can be explained as follows. When all components are operational, the lifetimes of all components are represented by a continuous cumulative distribution function (cdf) $F_1$. After the failure of one of the components at time $x_1>0$, the residual lifetimes of the remaining $n-1$ components are assumed to be equal in law to the residual lifetimes of components that already reached time $x_1$ with lifetime distribution represented by the cdf $F_2$. This process is repeated until the last failure time $x_s,\, s\leq n$ is observed. 

Throughout, we use the term `failure times' for the measurements. However, it should be mentioned that the SOS model is not limited to describe failure times but can also be used to model `time to event' occurrences, such as a liner wearing thin for example.

Most of the early work on SOS operated assumed a sufficiently large sample of failure times. This assumption was relaxed in \cite{Balakrishnan2011} and \cite{Balakrishnan2008}. The authors consider small sample sizes instead. The former by introducing different types of link functions between model parameters. The latter incorporates an order restriction on the underpinning model parameters $\alpha_i,\, i \in \calN$, defined by the CPHR. In the CPHR model, the underpinning distributions are given by
\begin{align*}
F_i = 1-\big(1-F\big)^{\alpha_{i}},\, i\in \calN,
\end{align*}
for some absolutely continuous baseline cdf $F$. The authors derive OR MLEs and demonstrate how these estimates perform better than their unrestricted counterparts when sample sizes are small. The justification of the simple order restriction originates from a conceptual idea that the failure of a component can only increase the risk of failure for the remaining components. The restriction $0\leq \alpha_1 \leq \dots \leq \alpha_n$ on the model parameters represents such an increasing risk of failure under the CPHR. On the other hand, unrestricted ML estimators can produce estimates that dissent from this restriction, especially if sample sizes are small. For this reason, the authors recommend using OR MLEs instead. Additional reading on ML estimation under order restriction can be found in \cite{Kim2004}.

The SOS model was also looked at from a Bayesian perspective. In \cite{Burkschat2010}, the authors apply independent order statistic priors on the model parameters to incorporate the assumption of increasingly ordered parameters. For further applications of Bayesian statistics, see \cite{Schenk2011,Shafay2014,Ahmadi2018,Tsai2021}. Many of these authors link the model of SOS to different types of censoring, but only the latter consider heterogeneous components in their analyses.

\cite{Baratnia2017} introduced the extension to heterogeneous components, dropping the assumption that components are identically distributed. They proposed the model of ESOS, which considers independent but possibly heterogeneous components instead. For more insights on heterogeneous components, we also refer the reader to \cite{Liu1998}, who considered system reliability in a load share accelerated testing design. A general discussion of models with heterogeneous components is provided by \cite{Balakrishnan2007c}. The ESOS extension requires a more complex notation to distinguish between the individual components. For this purpose, let the $n$ components be labelled consecutively from $1$ to $n$ and introduce indicator random variables $C_i,\,i\in\{1,\dots,n\}$ tracking the source of the $i$-th failure component. The random vector {$\Pi_k=(C_k,\dots,C_1),\,1\leq k\leq n$} then consists of the sequence of labels from failed components until the $k$-th failure in reversed order. For any given failure sequence $\pi_k=(c_k,\dots,c_1)$ let $B_{\pi_k}=\{1,\dots, n\}\setminus\{c_1,\dots, c_k\}$ be the set of component labels surviving the $k$-th failure.

Similar to the construction process of ordinary SOS, ESOS are generated as follows: at time $x_0=0$ all components are assumed to be operational, and the associated lifetime distributions are represented by $n$ many independent random variables $Y_1^{(1)}, \dots, Y_n^{(1)}$, where ${Y_i^{(1)}\sim F_i}, i\in \calN$. Then, the first failure is recorded as the minimum of those, i.e. $X_*^{(1)} = \min \{Y_1^{(1)}, \dots, Y_n^{(1)}\}$. Given that component $c_1\,\in\calN$ was the first to fail at time $X_*^{(1)}= x_1$, the residual lifetimes of the remaining $n-1$ components are assumed to be equal in law to the residual lifetimes of components that already reached time $x_1$ and are represented by independent random variables $Y_{i|c_1}^{(2)}=Y_{i|\pi_1}^{(2)},\, i\in B_{\pi_1}$  with distribution functions $Y_{i|\pi_1}^{(2)} \sim F_{i|\pi_1}$, where $F_{i|\pi_1}= F_{i}(\cdot\mid\pi_1), i\in B_{\pi_1}$, are supposed to be absolutely continuous distribution functions satisfying the technical restriction $F_{i|\pi_{1}}^{-1}(1)\leq F_{\ell|\pi_2}^{-1}(1),\,i\in B_{\pi_1},\,\ell\in B_{\pi_{2}}$. For each $Y_{i|\pi_1}^{(2)}$, construct random variables $X_{i|\pi_1}^{(2)}$ which realise the truncation at $X_*^{(1)}$ and based upon which the second failure time  ${X_{*}^{(2)} = \min \big\{X_{i|\pi_1}^{(2)}: i \in B_{\pi_1} \big\}}$ is defined. This process is repeated until the $s$-th failure is observed. 

\begin{figure}
\begin{centering}
\begin{tikzpicture}

\node [circle, draw, fill=black] (root) at (0,0) {};
\node (a1) at (-3.5,-1.5) {$F_1$};
\node [rectangle, draw] (a2) at (0,-1.5) {$F_2$};
\node (a3) at (3.5,-1.5) {$F_3$};
\node (b1) at (-4.5,-3) {$F_{2|1}$};
\node (b2) at (-3,-3) {$F_{3|1}$};
\node (b3) at (-1,-3) {$F_{1|2}$};
\node [rectangle, draw] (b4) at (1,-3) {$F_{3|2}$};
\node (b5) at (3,-3) {$F_{1|3}$};
\node (b6) at (4.5,-3) {$F_{2|3}$};
\node (c1) at (-4.5,-4.5) {$F_{3|2,1}$};
\node (c2) at (-3,-4.5) {$F_{2|3,1}$};
\node (c3) at (-1,-4.5) {$F_{3|1,2}$};
\node [rectangle, draw] (c4) at (1,-4.5) {$F_{1|3,2}$};
\node (c5) at (3,-4.5) {$F_{2|1,3}$};
\node (c6) at (4.5,-4.5) {$F_{1|2,3}$};

\draw (root) -- (a1) -- (b1) -- (c1);
\draw (a1) -- (b2)--(c2);
\draw (root) -- (a2) -- (b3)--(c3);
\draw (a2) -- (b4)--(c4);
\draw (root) -- (a3) -- (b5) -- (c5);
\draw (a3) -- (b6) -- (c6);

\node (root2) at (7,0)  {$x_0$};
\node (a10) at (7,-1.5)  {$x_1,\, c_1=2$};
\node (a11) at (7,-3)  {$x_2,\, c_2=3$};
\node (a12) at (7,-4.5)  {$x_3,\, c_3=1$};

\draw[->] (root2) -- (a10);
\draw[->] (a10) -- (a11);
\draw[->] (a11) -- (a12);

\draw[dashed] (5.5,0) -- (5.5,-4.7);

\end{tikzpicture}
\caption{Example of ESOS data generating process for $n=s=3$ with failure sequence $\pi_3=(1,3,2)$; rectangles highlight the estimable cdfs.}
\label{Fig:process}
\end{centering}
\end{figure}
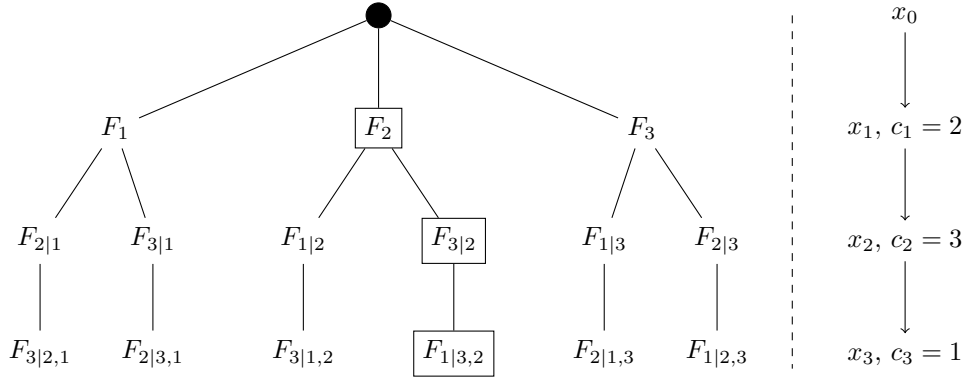

Figure \ref{Fig:process} illustrates all of the 15 distribution functions which are originally considered for a 3-out-of-3:F system. Only some of which can be estimated however, depending on the actual observed failure sequence.

For a detailed description of the construction process, see \cite{Pesch2023}. The formal definition yields that the random variables $X_{*}^{(1)},\dots,X_*^{(s)}$ with associated failure sources $C_1,\dots,C_s$ are called the first $s$ ESOS in a system of size $n$, if the joint density function is given by
\begin{align}
\label{eq-jointpdf}
f_{X_*^{(s)},C_s,\dots,X_*^{(1)},C_1}(x_s,c_s,\dots,x_1,c_1)= \prod_{k=1}^{s}\Bigg( \frac{f_{c_k|\pi_{k-1}}(x_{k})}{\overline{F}_{c_k|\pi_{k-1}}(x_{k-1})} \prod_{j\in B_{\pi_k}} \frac{\overline{F}_{j |\pi_{k-1}}(x_{k})}{\overline{F}_{j|\pi_{k-1}}(x_{k-1})}\Bigg),
\end{align}
$x_0 < x_1 < \dots < x_s$ and where $\overline{F}_{j}(x_0)=1,\, j \in \mathcal{N}$. 
The authors also derive the MLEs of the underpinning distribution parameters in the case of exponentially distributed lifetimes. While these are shown to perform well for reasonably large samples, the problem of counterintuitively ordered estimates emerges for small sample sizes. The following example illustrate this further.

\begin{example}
Consider a system with $n=3$ components that is observed until the second failure; lifetimes are supposed to be exponentially distributed. Assume that component 1 fails first. The risk of failure for the other two components should then increase compared to before. The associated distribution parameters can express this increase. In the exponential case, we would expect $\lambda_{2|1} > \lambda_{2}$ for example, where $\lambda_{2|1}$ describes the risk of failure for component 2 after the failure of component 1 and $\lambda_{2}$ describes the risk of failure for component 2 when all components are still operational. However, the associated MLEs may not adhere to this order and may yield $\widehat{\lambda}_{2}>\widehat{\lambda}_{2|1}$ instead and therefore suggest a counterintuitive attribution of risks.
\end{example}

Our aim is to address this issue and derive OR MLEs, which respect the natural order of the model parameters. These estimators will be shown to perform better than their ordinary counterparts, especially when sample sizes are small.
 
\subsection{ESOS under the CPHR model}
\label{Sec-1}

Similarly to \cite{Cramer1996} a particular choice of underpinning distribution functions is considered.

\begin{Ass}
\label{Ass-CPHR}
In the following, consider component distributions 
\begin{align*}
    F_{j|\pi_{k-1}} = 1-(1-F_j^*)^{\alpha_{j|\pi_{k-1}}},\, j \in \{1,\dots,n\},
\end{align*}
with some absolutely continuous and strictly increasing baseline distribution function $F_j^*$, which is referred to as the baseline cdf of the $j$-th component.
\end{Ass}

Due to Assumption \ref{Ass-CPHR}, the density function $f_{j|\pi_{k-1}}$ can be expressed based on the baseline cdfs and pdfs.
\begin{equation}
\label{eq-pdfCPHR}
f_{j|\pi_{k-1}} = \alpha_{j|\pi_{k-1}} (1-F^*_{j})^{\alpha_{j|\pi_{k-1}}-1} f_j^*
\end{equation}
This leads to the hazard rate functions

\begin{align}
\label{eq-Hazard}
    h_{j|\pi_{k-1}}= \frac{f_{j|\pi_{k-1}}}{\overline{F}_{j|\pi_{k-1}}} = \alpha_{j|\pi_{k-1}} \,h^*_j,
\end{align}

where $h_j^*$ is the baseline hazard rate function of component $j$, i.e. $h_j^*=f_j^*/\overline{F}_j^*$.

In the CPHR model, the shape of the baseline function uniquely determines the shape of the corresponding cdfs $F_{j|\pi_{k-1}}$. If the baseline distribution is an exponential distribution with parameter $\lambda$, the corresponding cdfs $F_{j|\pi_{k-1}},\, j\in \calN$, representing the different stages for component $j$, are also exponentially distributed with parameters $\lambda \, \alpha_{j|\pi_{k-1}}$. If the baseline distribution is a Weibull distribution, then the corresponding cdfs $F_{j|\pi_{k-1}}$ are also Weibull distribution functions.

Under Assumption \ref{Ass-CPHR} and due to \eqref{eq-pdfCPHR}, the joint density of the first $s$ ESOS as presented in \eqref{eq-jointpdf} can be written as
\begin{align}
\label{eq-jointpdfCPHR}
f_{X_*^{(s)},C_s,\dots,X_*^{(1)},C_1}(x_s,c_s,\dots,x_1,c_1)
&= \prod_{k=1}^{s}\Bigg(\alpha_{c_k|\pi_{k-1}}\frac{f_{c_k}^*(x_k)}{\overline{F}_{c_k}^*(x_k)}\bigg( \frac{\overline{F}_{c_k}^*(x_{k})}{\overline{F}_{c_k}^*(x_{k-1})}\bigg)^{\alpha_{c_k|\pi_{k-1}}}\notag \\
& \quad \times \prod_{j\in B_{\pi_k}}\bigg( \frac{\overline{F}_j^*(x_{k})}{\overline{F}_j^*(x_{k-1})}\bigg)^{\alpha_{j|\pi_{k-1}}}\Bigg),
\end{align}
$0 = x_0 < x_1 < \dots < x_s$ and where $\overline{F}^*_{j}(x_0)=1,\, j \in \mathcal{N}$. 

\begin{Rem}
Representation \eqref{eq-jointpdfCPHR} illustrates the relation of the model to the SOS model. In fact, the joint density of the first $s$ SOS as displayed in \cite{Cramer1996} (also refer to \eqref{eq:SOS}) can be obtained from \eqref{eq-jointpdfCPHR} if we replace $\alpha_{c_k|\pi_{k-1}}$ with $\alpha_k$ and the baseline distributions $f_{c_k}^*$ and $F_{c_k}^*$ with $f$ and $F$ respectively. Note that representation \eqref{eq-jointpdfCPHR} involves the information about failed components in the model. By construction this information is not included in the standard SOS model. If we were to consider the failure sequence in the SOS model, and with the identifications from above, representation \eqref{eq-jointpdfCPHR} would become independent of $c_1,\dots,c_s$. This indicates a discrete uniform distribution of the failure sequences $\Pi_k,\,1\leq k \leq s$. To be more precise: $\Pi_k$ is uniformly distributed with support $\mathfrak{S}_{n,k}$ in that case, i.e. $P(\Pi_k=\pi_k)=(n-k)!/n!,\, k \in \{1,\dots,n\}$, where $\mathfrak{S}_{n,k}$ denotes the $k$-permutations from the set $\{1,\dots,n\}$. Therefore, knowledge of the failure sequence $\Pi_s$ in the SOS model would not yield any additional information. In conclusion, this justifies the omission of the use of the failure sequence in the SOS model. 
\end{Rem}

\begin{figure}
\begin{centering}
\begin{tikzpicture}

\node [circle, draw, fill=black] (root) at (0,0) {};
\node [rectangle, draw] (a1) at (-3.5,-1.5) {$F_{1}$};
\node [rectangle, draw] (a2) at (0,-1.5) {$F_{1}$};
\node [rectangle, draw] (a3) at (3.5,-1.5) {$F_{1}$};
\node [rectangle, draw] (b1) at (-4.5,-3) {$F_{2}$};
\node [rectangle, draw] (b2) at (-3,-3) {$F_{2}$};
\node [rectangle, draw] (b3) at (-1,-3) {$F_{2}$};
\node [rectangle, draw] (b4) at (1,-3) {$F_{2}$};
\node [rectangle, draw] (b5) at (3,-3) {$F_{2}$};
\node [rectangle, draw] (b6) at (4.5,-3) {$F_{2}$};
\node [rectangle, draw] (c1) at (-4.5,-4.5) {$F_{3}$};
\node [rectangle, draw] (c2) at (-3,-4.5) {$F_{3}$};
\node [rectangle, draw] (c3) at (-1,-4.5) {$F_{3}$};
\node [rectangle, draw] (c4) at (1,-4.5) {$F_{3}$};
\node [rectangle, draw] (c5) at (3,-4.5) {$F_{3}$};
\node [rectangle, draw] (c6) at (4.5,-4.5) {$F_{3}$};

\draw (root) -- (a1) -- (b1) -- (c1);
\draw (a1) -- (b2)--(c2);
\draw (root) -- (a2) -- (b3)--(c3);
\draw (a2) -- (b4)--(c4);
\draw (root) -- (a3) -- (b5) -- (c5);
\draw (a3) -- (b6) -- (c6);

\node (root2) at (7,0)  {$x_0$};
\node (a10) at (7,-1.5)  {$x_1,\, c_1=2$};
\node (a11) at (7,-3)  {$x_2,\, c_2=3$};
\node (a12) at (7,-4.5)  {$x_3,\, c_3=1$};

\draw[->] (root2) -- (a10);
\draw[->] (a10) -- (a11);
\draw[->] (a11) -- (a12);

\draw[dashed] (5.5,0) -- (5.5,-4.7);

\end{tikzpicture}
\caption{Example of SOS data generating process for $n=s=3$ with failure sequence $\pi_3=(1,3,2)$; rectangles highlight the estimable cdfs.}
\label{Fig:SOS}
\end{centering}
\end{figure}
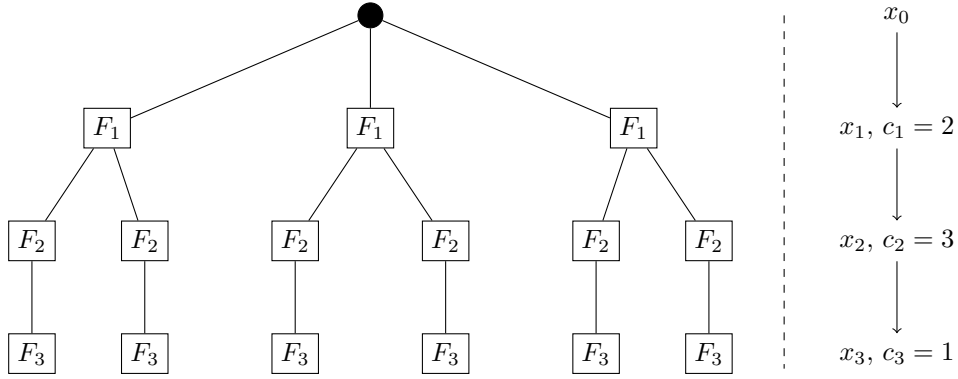

In contrast to Figure \ref{Fig:process} and the described ESOS model, Figure \ref{Fig:SOS} showcases, that regardless of which component fails, the same cdf is observed on the respective level in the SOS model. The outlined relation between the two models also illustrates the role of the indicator variables $C_i$, which govern the choice of distributions. In contrast to SOS, the ESOS model considers multiple distribution functions on each level, but ultimately only one (depending on the source of failure on that level) is estimable. The remaining distributions cannot be estimated as no MLEs of the associated parameters exists due to the monotonicity of the likelihood function (refer to \eqref{eq-LL}). 

The unrestricted MLEs for the observed cdfs can be readily obtained. 

\begin{Lemma}
\label{Lem-unreMLE}
If the baseline distributions $F_j^*$ are known, then the unrestricted MLEs of $\alpha_{j|\pi_{k-1}},\, j\in B_{\pi_{k-1}}$ exist and are given by
\[ \widehat{\alpha}_{j|\pi_{k-1}} =
  \begin{cases}
    \frac{m_{j|\pi_{k-1}}}{\sum_{i=1}^r \delta_{j,k,i} \mathds{1}_{\{\pi_{k-1}\}}(\pi_{k-1,i})}, & (j,\pi_{k-1})\text{ observed} \\
    0, & \text{ else}
\end{cases} \]
where $\delta_{j,k,i}=\ln(\overline{F}_j^* (x_{k-1,i})) - \ln(\overline{F}_j^* (x_{k,i})),\,\mathds{1}_{\{\pi_{0}\}} (\pi_{0,i})= 1,\,i \in \{1,\dots,r\}$ and $m_{j|\pi_{k-1}}$ counts how often the failure sequence $(j,\pi_{k-1})$ was observed within the $r$ trials, i.e.:
\[ m_{j|\pi_{k-1}} = \sum_{l=1}^{r} \mathds{1}_{\{c_1,\dots, c_{k-1},j\}}(c_{1,l},\dots,c_{k-1,l},c_{k,l}). \]
\end{Lemma}

\textit{Proof}: 

For $r$ samples of the first $s$ ESOS including failure sources $\big(X_{*i}^{(1)},C_{1,i},X_{*i}^{(2)},C_{2,i},\dots,X_{*i}^{(s)},C_{s,i}\big)_{1\leq i \leq r}$ , the log-likelihood function reads
\begin{align}
\label{eq-LL}
    l(\boldsymbol{\alpha}) = \sum_{i=1}^r \sum_{k=1}^s \Bigg(\ln\big(\alpha_{c_{k}|\pi_{k-1,i}}\big) + \ln\bigg(\frac{f^*_{c_{k,i}}(x_{k,i})}{\overline{F}^*_{c_{k,i}}(x_{k,i})}\bigg) + \sum_{j\in B_{\pi_{k-1,i}}}  \alpha_{j|\pi_{k-1,i}} \,\ln \bigg(\frac{\overline{F}^*_{j}(x_{k,i})}{\overline{F}^*_{j}(x_{k-1,i})}\bigg)\Bigg).
\end{align}

For any failure sequence $(j,\pi_{k-1})$, the structure of $l(\boldsymbol{\alpha})$ is such, that it can be additively separated, into $l(\boldsymbol{\alpha})= K + l^*(\alpha_{j|\pi_{k-1}})$, where only the latter term depends on the parameter $\alpha_{j|\pi_{k-1}}$. Thus, we get
\begin{align*}
l^*(\alpha_{j|\pi_{k-1}}) = m_{j|\pi_{k-1}}\,\ln(\alpha_{j|\pi_{k-1}}) + \alpha_{j|\pi_{k-1}} \sum_{i=1}^r \delta_{j,k,i}\mathds{1}_{\{\pi_{k-1}\}} (\pi_{k-1,i}),
\end{align*}
which can be maximised for level 1 parameters via
\begin{align*}
    \frac{\partial  l}{\partial \alpha_j} (\boldsymbol{\alpha}) = \frac{m_j}{\alpha_j} + \sum_{i=1}^r \ln\big(\overline{F}_j^*(x_{1,i})\big) 
\end{align*}
and for all other levels via
\begin{align*}
    &\frac{\partial  l}{\partial \alpha_{j|\pi_{k-1}}} (\boldsymbol{\alpha}) =  \frac{m_{j|\pi_{k-1}}}{\alpha_{j|\pi_{k-1}}} + \sum_{i=1}^r \delta_{j,k,i}
    \mathds{1}_{\{\pi_{k-1}\}} (\pi_{k-1,i}).
\end{align*}

Equating the above to zero and rearranging leads to an expression for the ML estimates. Using the inequality $\ln t\le t-1$, $t>0$ (with equality iff $t=1$), we get the upper bound:
\begin{align*}
l^*(\alpha_{j|\pi_{k-1}}) &= m_{j|\pi_{k-1}}\,\ln(\alpha_{j|\pi_{k-1}})  - m_{j|\pi_{k-1}}\, \frac{\alpha_{j|\pi_{k-1}}}{\widehat{\alpha}_{j|\pi_{k-1}}}\\
&\leq m_{j|\pi_{k-1}} \big( \ln(\widehat{\alpha}_{j|\pi_{k-1}}) - 1\big)\\
&= l^*(\widehat{\alpha}_{j|\pi_{k-1}}),
\end{align*}
where equality holds if $\alpha_{j|\pi_{k-1}}=\widehat{\alpha}_{j|\pi_{k-1}}$. This proves that $\widehat{\alpha}_{j|\pi_{k-1}}$ 
is indeed the MLE of $\alpha_{j|\pi_{k-1}}$, provided that the failure sequence $(j,\pi_{k-1})$ is observed at least once.

\begin{flushright}
$\square$
\end{flushright}

\begin{Rem}
\label{Rem-Structure}
The structure of the proposed MLEs is similar to those based on SOS (see \cite{Cramer1996}). However, in contrast to the MLEs based on SOS, $\widehat{\alpha}_{j|\pi_{k-1}}$ and $\widehat{\alpha}_{j,k}$ (refer to \eqref{eq-MLEunre}) do not only depend on the failure times, but additionally on how often a certain sequence of failures was observed. These failure sequences are random as they are represented by the $C_i$. Again, this illustrates the difference of the two models, which is caused by the introduction of heterogeneous components.
\end{Rem}

\subsection{History independent load change}
\label{Sec-2}

\begin{figure}[ht]
\begin{centering}
\begin{tikzpicture}

\node [circle, draw, fill=black] (root) at (0,0) {};
\node (a1) at (-3.5,-1.5) {$F_{1,1}$};
\node [rectangle, draw] (a2) at (0,-1.5) {$F_{2,1}$};
\node (a3) at (3.5,-1.5) {$F_{3,1}$};
\node (b1) at (-4.5,-3) {$F_{2,2}$};
\node [rectangle, draw] (b2) at (-3,-3) {$F_{3,2}$};
\node (b3) at (-1,-3) {$F_{1,2}$};
\node [rectangle, draw] (b4) at (1,-3) {$F_{3,2}$};
\node (b5) at (3,-3) {$F_{1,2}$};
\node (b6) at (4.5,-3) {$F_{2,2}$};
\node (c1) at (-4.5,-4.5) {$F_{3,3}$};
\node (c2) at (-3,-4.5) {$F_{2,3}$};
\node (c3) at (-1,-4.5) {$F_{3,3}$};
\node [rectangle, draw] (c4) at (1,-4.5) {$F_{1,3}$};
\node (c5) at (3,-4.5) {$F_{2,3}$};
\node [rectangle, draw] (c6) at (4.5,-4.5) {$F_{1,3}$};

\draw (root) -- (a1) -- (b1) -- (c1);
\draw (a1) -- (b2)--(c2);
\draw (root) -- (a2) -- (b3)--(c3);
\draw (a2) -- (b4)--(c4);
\draw (root) -- (a3) -- (b5) -- (c5);
\draw (a3) -- (b6) -- (c6);

\node (root2) at (7,0)  {$x_0$};
\node (a10) at (7,-1.5)  {$x_1,\, c_1=2$};
\node (a11) at (7,-3)  {$x_2,\, c_2=3$};
\node (a12) at (7,-4.5)  {$x_3,\, c_3=1$};

\draw[->] (root2) -- (a10);
\draw[->] (a10) -- (a11);
\draw[->] (a11) -- (a12);

\draw[dashed] (5.5,0) -- (5.5,-4.7);

\end{tikzpicture}
\caption{Example of ESOS data generating process for $n=s=3$ under history independence; rectangles highlight the estimable cdfs.}
\label{Fig:process2}
\end{centering}
\end{figure}
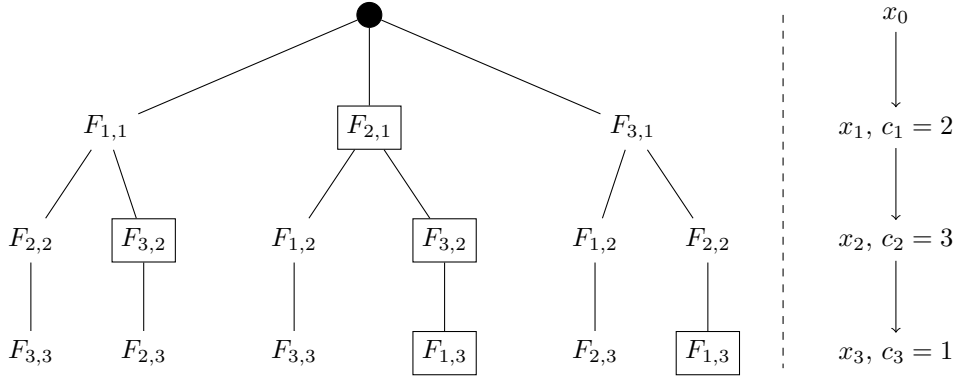

\begin{Ass}
\label{Ass-EQUALLOAD}
Consider the situation where neither the exact sequence of failure nor the set of failed components matters and that each component failure affects the surviving components in exactly the same way. In that sense, the failure history of a system is irrelevant. The load change, represented in the change of lifetime distributions, solely depends on the number of failed components. To this end, the following identifications of model parameters are made:
\begin{align*}
    \alpha_{j|\pi_{k-1}} = \alpha_{j,k}, \quad j\in B_{\pi_{k-1}},\, k \in \{1,\dots,s\} ,
\end{align*}
where $k$ denotes the level depth. This leads to the corresponding component cdfs
\begin{align*}
    F_{j,k} = 1-(1-F_j^*)^{\alpha_{j,k}}, 
\end{align*}
for component $j\in \{1,\dots,n\}$ on level $k \in \{1,\dots,s\}$ with baseline distribution $F_j^*$.
For brevity, we will refer to this assumption as `history independence' moving forward.  Figure \ref{Fig:process2} summarises the situation.
\end{Ass}

\begin{Lemma}
\label{Lemma-Alpha}
Under Assumption \ref{Ass-EQUALLOAD} and if the baseline cdfs $F_j^*,\,j\in \calN$ are known, the unrestricted MLEs exist and are given by
\begin{align}
\label{eq-MLEunre}
\widehat{\alpha}_{j,k} =\begin{cases}
    \frac{m_{j,k}}{\sum_{i=1}^r \delta_{j,k,i} (I_{j,k-1})_i,   }& m_{j,k}>0 \\
    0, & m_{j,k}=0
\end{cases},
\end{align}
where $m_{j,k}$ count how often component $j$ failed on level $k$ within the $r$ trials and $I_{j,k}$ indicates whether component $j$ has not failed on the first $k$ levels, i.e. it was still functioning on level $k+1$. Therefore, we consider the random counters
\begin{align*}
    M_{j,k} = \sum_{i=1}^r \mathds{1}_{\{j\}} (C_{k,i})
\end{align*}
and
\begin{align*}
    I_{j,k}= 1 - \sum_{i=1}^k \mathds{1}_{\{j\}}(C_i)
\end{align*}
respectively. Finally, by convention, let $I_{j,0}=1$, $j\in \calN$.
\end{Lemma}

\textit{Proof}:

The proof is along the lines of the proof of Lemma \ref{Lem-unreMLE} with representation \eqref{eq-jointpdfCPHR} simplifying to
\begin{align*}
f_{X_*^{(s)},C_s,\dots,X_*^{(1)},C_1}(x_s,c_s,\dots,x_1,c_1)= \prod_{k=1}^{s}\alpha_{c_k,k}\,\frac{f_{c_k}^*(x_k)}{\overline{F}_{c_k}^*(x_k)} \prod_{j\in B_{\pi_{k-1}}}\bigg( \frac{\overline{F}_j^*(x_{k})}{\overline{F}_j^*(x_{k-1})}\bigg)^{\alpha_{j,k}}.
\end{align*}
Optimisation of the likelihood function 
\begin{align}
\label{eq-LLF}
L(\boldsymbol{\alpha}) &= K \,\prod_{k=1}^s\, \prod_{j\in \mathcal{N}} \alpha_{j,k}^{m_{j,k}}\, \prod_{i=1}^r \, \prod_{k=1}^s \, \prod_{j\in B_{\pi_{k-1,i}}} e^{-\alpha_{j,k} \delta_{j,k,i}}\notag\\
&=  K \,\prod_{k=1}^s\, \prod_{j\in \mathcal{N}} \alpha_{j,k}^{m_{j,k}}\,\prod_{i=1}^r\, \prod_{k=1}^s\, \prod_{j\in \mathcal{N}}\,e^{-\alpha_{j,k} \delta_{j,k,i} (I_{j,k-1})_i}
\end{align}
where $K=\prod_{i=1}^r\,\prod_{k=1}^s\,\frac{f_{c_{k,i}}^*(x_{k,i})}{\overline{F}^*_{c_{k,i}}(x_{k,i})}$ is independent of $\boldsymbol{\alpha}$, yields the MLEs (if $m_{j,k}>0$).

\begin{flushright}
$\square$
\end{flushright}

\begin{Rem}
As outlined in Remark \ref{Rem-Structure}, the existence of the MLEs depends on the observed sequences of failures. Simultaneously, the dimension of the parameter space increases exponentially in the system size $n$. Consequently, the larger a system and the smaller the sample size, the smaller is the proportion of existing estimates. Refer to Table \ref{Table:Mean} for an example. In this context, Assumptions \ref{Ass-CPHR} and \ref{Ass-EQUALLOAD} are not only reasonable for many real life applications but additionally support the goal of reducing the dimension of the parameter space. In Figure \ref{Fig:process2} we consider the same case as in Figure \ref{Fig:process} but only 9 instead of 15 distributions are required and the proportion of estimable cdfs is increased.
\end{Rem}

\section{Order Restricted MLEs from ESOS}
\label{Sec-3}

All examples and results in this section operate under the CPHR and under the history independence assumption, i.e. Assumptions \ref{Ass-CPHR} and \ref{Ass-EQUALLOAD}.

The proposed MLEs from the previous two sections are valuable and meaningful in situations where they yield feasible estimates. However, in some cases they may lead to estimates which do not align with the expectation that failure of one component increases the stress on the remaining components and thereby increases their risk of failure. The following, simulated data example based on failure times and sources from $r=10$ many systems illustrates the problem. All systems are considered to feature three components and were observed until second failure.

\begin{table}[ht]
\centering
\caption{Data example of failure times and sources for a 2-out-of-3 system.}
\label{table:1}
\begin{tabular}{c|cccccccccc}
\hline
 $ l $ &  $1 $  &  $2 $  &  $ 3 $  &  $ 4 $ &  $5 $&  $6 $&  $7 $&  $8 $&  $9 $&  $10 $ \\
\hline
$x_1$ & 0.22 & 0.01 &0.23 &0.14 &0.24 &0.05 &0.17 &0.37 &0.05&
0.16 \\
$c_1$ & 1  & 1  & 3  & 1 & 2  & 1  & 2 & 1 & 1  & 2 \\
\hline
$x_2$ & 0.58 & 0.32& 0.84 & 0.32& 0.39& 0.20& 0.25& 1.32 & 0.29 & 0.21 \\
$c_2$ &2  & 2  & 2  & 3 & 1  & 2  & 1 & 3& 2  & 1 \\
\hline
\end{tabular}
\end{table}



\begin{example}
\label{ex-ORvio}
Considering exponential baseline distributions with rate parameter $\lambda=1$ for all three components, the following MLEs based on Lemma \ref{Lemma-Alpha} for component 3 are derived: $\widehat{\alpha}_{3,1}=0.61 < 0.82=\widehat{\alpha}_{3,2}$. These estimates are in line with the idea of component failure having a negative effect on the lifetime expectation of the remaining components, in the sense that the risk of failure for the remaining components increases. The estimates for component 1 however, are $\widehat{\alpha}_{1,1}=3.65 >3.34=\widehat{\alpha}_{1,2}$, which indicates that the risk of failure for this component has reduced after one of the other two components has failed. This is in clear conflict with the belief that the risk of failure should never decrease when components fail sequentially. 
\end{example}

To adequately model the increase in risk of failure and therein address the illustrated problem, the following assumption is made.

\begin{Ass}
\label{ass:order}
Assume a simple order restriction
\begin{align*}
    \alpha_{j,k} \geq \alpha_{j,l}\,,\,  k\geq l \in \{1,\dots,s\},\,  j\in \{1,\dots,n\}.
\end{align*}
of the underpinning model parameters. 
\end{Ass}

The simple order restriction in Assumption~\ref{ass:order} on the model parameters directly implies a stochastic ordering of the associated lifetime distributions $F_{j,k}$ w.r.t.\ the hazard rate order (see \cite[Chapter 3]{Navarro2021}), that is, the hazard rates $h_{j,k}$ of the lifetimes are ordered. This ordering is intuitive in the sense that the hazard rate of any component should increase as components fail successively, i. e. across levels.

\begin{Lemma}
\label{Lemma-HR} For arbitrary but fixed $k\geq l \in \{1,\dots,s\}$ with  $\alpha_{j,k}\geq \alpha_{j,l}$, we obtain:
\[
X_{j,l}^{(k+1)} \preceq_{hr} X_{j,k}^{(l+1)},\quad j\in \{1.\dots,n\},
\]
where $X_{j,k}^{(k+1)}$ and $X_{j,l}^{(l+1)}$ are defined analogously to $X_{i|\pi_k}^{(k+1)}$ from Section \ref{Sec-0} but under history independence, and $\preceq_{hr}$ denotes the hazard rate order.
\end{Lemma}

\textit{Proof}: 

Under the history independence assumption representation \eqref{eq-Hazard} simplifies to
\[
    h_{j,k}(x) = \frac{f_{j,k}(x)}{\overline{F}_{j,k}(x)} = \alpha_{j,k} \,h^*_{j} (x),
\]
where $h^*_{j}$ is the baseline hazard rate of component $j$. For $\alpha_{j,k}\geq \alpha_{j,l}$, clearly
\[
h_{j,l}(x)=\alpha_{j,l}\,h^*_{j}(x) \leq \alpha_{j,k}\,h^*_{j}(x) = h_{j,k}(x), \quad x\geq 0.
\]
This proves the stochastic order.

\begin{flushright}
$\square$
\end{flushright}

\bigskip

We now present the new set of estimates and thereby the main result. 

\begin{Theo}
\label{Theo-1}
If the baseline cdfs $F_j^*,\,j\in \{1,\dots,n\}$ are known and the unrestricted MLEs $\widehat{\alpha}_{j,k}$ exist, then the MLEs under simple order restriction $\widehat{\alpha}_{j,k}^*$ also exist. In this case, they are given as the reciprocal of the solution of the isotonic regression over the inversed unrestricted MLEs, i.e.,
\begin{align*}
\widehat{\alpha}_{j,s-k+1}^* = \bigg(\max_{l \leq k} \min_{t \geq k} \frac{\sum_{\nu=l}^t \widehat{\alpha}_{j,s-\nu +1}^{-1} m_{j,s-\nu+1}}{\sum_{\nu=l}^t m_{j,s-\nu+1}}\bigg)^{-1}, \quad 1 \leq k \leq s. 
\end{align*}

\end{Theo}

\textit{Proof}: 

Firstly note that the likelihood function \eqref{eq-LLF} can also be expressed as
\begin{align}
\label{eq-SEVEN}
    L(\boldsymbol{\alpha}) = K \,\prod_{k=1}^s\, \prod_{j\in \mathcal{N}} \alpha_{j,k}^{m_{j,k}}\,\, \prod_{i=1}^r \, \prod_{k=1}^s \, \prod_{j\in \mathcal{N}} e^{-\alpha_{j,k} \delta_{j,k,i}\, (I_{j,k-1})_i}.
\end{align}

Due to its structure, it is obvious that all $\alpha_{j,k}$ can be separated multiplicatively from another. Therefore, it suffices to perform ML estimation only for a single component, say component $\ell \in \{1,\dots,n\}$. With respect to $\boldsymbol{\alpha}_{\ell}=(\alpha_{\ell,1},\dots,\alpha_{\ell,s})$, the log-likelihood function can be expressed as ${l(\boldsymbol{\alpha})=l(\boldsymbol{\alpha}_{\ell}) + \widetilde{K}}$, where $\widetilde{K}$ is independent of $\boldsymbol{\alpha}_{\ell}$ and

\begin{align*}
l(\boldsymbol{\alpha}_{\ell}) &= \sum_{k=1}^s\, m_{\ell,k} \ln{(\alpha_{\ell,k})}\,\, \sum_{i=1}^r \, \sum_{k=1}^s -\alpha_{\ell,k} \delta_{j,k,i} (I_{\ell,k-1})_i\Big\} \\
&= \sum_{k=1}^s \Bigg\{ \bigg( \ln(\alpha_{\ell,k}) - \frac{\alpha_{\ell,k}}{m_{\ell,k}} \, \sum_{i=1}^r  \delta_{j,k,i} (I_{\ell,k-1})_i\Big\} \bigg)m_{c,k}\Bigg\}.
\end{align*}
From \eqref{eq-MLEunre},
\begin{equation}
\label{eq-LLFonecomp}
l(\boldsymbol{\alpha}_{\ell}) = \sum_{k=1}^s \Bigg\{ \bigg( \ln(\alpha_{\ell,k}) - \frac{\alpha_{\ell,k}}{ \widehat{\alpha}_{\ell,k}} \bigg)m_{\ell,k}\Bigg\},
\end{equation}
where $\widehat{\alpha}_{\ell,k}$ are the unrestricted MLEs from Lemma \ref{Lemma-Alpha}.
Consider the injective reparameterisation $\alpha_{\ell,k}= \lambda_{\ell,k}^{-1}$, as well as a reversed order of summation. Then, 

\begin{align}
\label{eq-ACHT}
    &l(\boldsymbol{\lambda}_{\ell}) = \sum_{k=1}^s \bigg( -\ln(\lambda_{\ell,k})m_{\ell,k} - \frac{m_{\ell,k}}{\lambda_{\ell,k}\,\widehat{\alpha}_{\ell,k}}  \bigg)\notag\\
    &= \sum_{k=1}^s  \bigg(-\ln(\lambda_{\ell,s-k+1})m_{\ell,s-k+1} 
    - \frac{m_{\ell,s-k+1}}{\lambda_{\ell,s-k+1}\,\widehat{\alpha}_{\ell,s-k+1}} \bigg).
\end{align}

The remainder of the proof follows the idea of Remark 3.3 in \cite{Balakrishnan2008}. To this end, further consider the convex function $\phi(u)=-\ln(u)$, such that based on equation (1.29) in \cite{barlow1972}, p. 39, the maximisation of \eqref{eq-ACHT} with respect to $\lambda_{\ell,s-k+1}$ and the choices $f_\ell(k)= \lambda_{\ell,s-k+1}$, $g_\ell(k)= \widehat{\alpha}_{\ell,s-k-1}^{-1}$ and $w_\ell(k)=m_{\ell,s-k+1}$ is equivalent to the minimisation of 
\begin{align*}
    \sum_{k=1}^s &\triangle_{\phi}\big(g_\ell(k),f_\ell(k)\big)w_\ell(k)\\
    &= \sum_{k=1}^s \bigg(\ln(f_\ell(k))+ \frac{g_\ell(k)}{f_\ell(k)}\bigg) w_\ell(k).
\end{align*}
Since $\alpha_{\ell,1}\leq \alpha_{\ell,2}\leq \dots \leq \alpha_{\ell,s}$, $\ell \in \{1,\dots,n\}$, $f_\ell(k)$ is non-decreasing in $k\in\{1,\dots,s\}$. Hence by Theorem 1.10 in \cite{barlow1972}, the isotonic regression
\begin{align*}
    g_\ell^*(k) = \widehat{\lambda}_{\ell,s-k+1}^* &= \max_{l\leq k} \min_{t \geq k} \frac{\sum_{\nu=l}^{t} g_\ell(\nu) \, w_\ell(\nu)}{\sum_{\nu=l}^{t} w_\ell(\nu) }= \max_{l \leq k} \min_{t \geq k} \frac{\sum_{\nu=l}^t \widehat{\alpha}_{j,s-\nu +1}^{-1} m_{j,s-\nu+1}}{\sum_{\nu=l}^t m_{j,s-\nu+1}}\\
    & = \max_{l\leq k} \min_{t \geq k} \frac{\sum_{\nu=l}^{t} \, \sum_{i=1}^r \delta_{\ell,s-\nu+1,i}\, (I_{\ell,s-\nu})_i}{\sum_{\nu=l}^{t} m_{\ell,s-\nu+1}}
\end{align*}
solves the above minimisation problem and subsequently yields MLEs $\widehat{\lambda}_{\ell,s-k+1}^*$, $k\in \{1,\dots,s\}$. Since MLEs are invariant under injective transformation, the final representation for the sought after MLEs $\widehat{\alpha}_{\ell,s-k+1}^*$, $k\in \{1,\dots,s\}$ under simple order restriction emerges.

\begin{flushright}
$\square$
\end{flushright}

\section{Simulation Studies}
\label{Sec-4}

Some extensive simulation studies are presented to explore the characteristics and describe the performance of the newly proposed estimators.

\subsection{Comparison of OR MLEs with ordinary MLEs}
\label{Sec-4.1}

\begin{example}
\label{Ex:1}
Consider a 3-out-of-4:F system that is observed $r=8$ many times. We computed the OR MLEs and their unrestricted counterparts for one of the heterogeneous components (say component 1) based on simulated data. The following results were achieved with true parameter choice $\alpha_{1,1}=2,\, \alpha_{1,2}=2.5$ and $\alpha_{1,3}=2.75$, where all underlying baseline distributions are assumed to be exponential distributions with rate parameter $\lambda=1$, $j \in\{1,\dots,n\}$, i.e., with density function $f_{j}(x)= e^{-x}\,\mathds{1}_{(0,\infty)}(x)$, $x\in\R$.
\end{example}

\begin{table}[ht]
\centering
\caption{Comparison of unrestricted and OR MLEs for a 3-out-of-4:F system based on $r=8$ observations with true model parameters $\alpha_{1,1}=2,\, \alpha_{1,2}=2.5$ and $\alpha_{1,3}=2.75$.}
\label{tab:FIRST}
\begin{tabular}{c|ccc|ccc}
case & $\widehat{\alpha}_{1,1}$ & $\widehat{\alpha}_{1,2}$ & $\widehat{\alpha}_{1,3}$ & $\widehat{\alpha}_{1,1}^*$ & $\widehat{\alpha}_{1,2}^*$ & $\widehat{\alpha}_{1,3}^*$ \\ \hline
1 & 3.37 & 0.85 & 4.87 & 1.93 & 1.93 & 4.87 \\
2 & 2.92 & 1.67 & 1.73 & 1.97 & 1.97 & 1.97 \\
3 & 5.05 & 2.41 & 1.97 & 3.66 & 3.66 & 3.66 \\
4 & 4.03 & 0.79 & 3.27 & 2.21 & 2.21 & 3.27 \\
5 & 2.19 & 3.01 & 2.96 & 2.19 & 2.99 & 2.99 \\
6 & 1.06 & 2.41 & 0.64 & 1.03 & 1.03 & 1.03
\end{tabular}
\end{table}

\begin{figure}[ht]
\begin{centering}
\includegraphics[width=85mm]{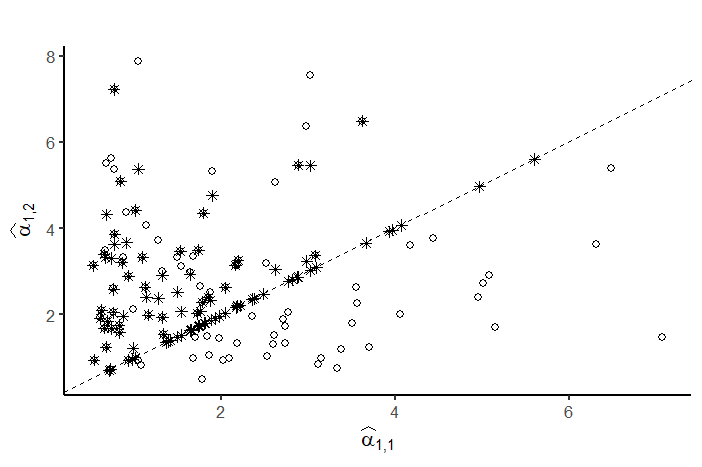}
\caption{Unrestricted ($\circ$) and OR ($\ast$) MLEs for level 1 and 2 model parameters}
\label{fig:1a}
\end{centering}
\end{figure}

\begin{figure}[ht]
\begin{centering}
\includegraphics[width=85mm]{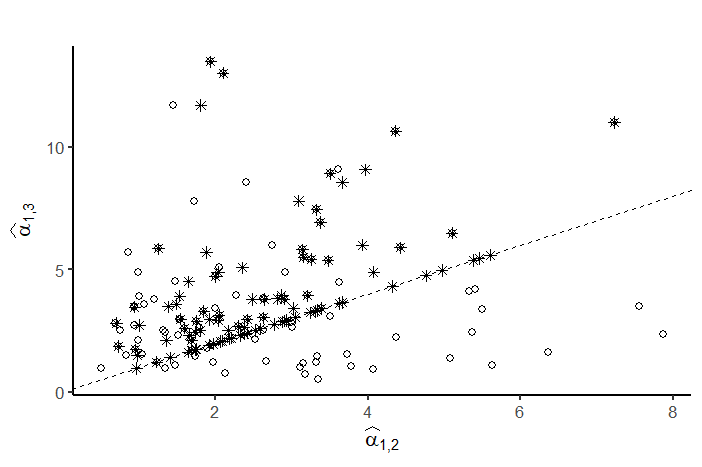}
\caption{Unrestricted ($\circ$) and OR ($\ast$) MLEs for level 2 and 3 model parameters}
\label{fig:1b}
\end{centering}
\end{figure}

Table \ref{tab:FIRST} illustrates the usefulness of the OR MLEs based on six test cases in which the unrestricted MLEs yield nonsensical results. The OR MLEs pose better estimates as they respect the order restriction. In both cases only $r=8$ systems were simulated. To further illustrate the generality of the results, Figures \ref{fig:1a} and \ref{fig:1b} depict unrestricted MLEs ($\circ$ symbol) from 100 simulations of a 3-out-of-4:F system as described in Example \ref{Ex:1} along with the corresponding OR MLEs ($\ast$ symbol) where they were well defined.

Taken as a vector, the triplet of unrestricted MLEs is projected onto the cone defined by $\alpha_{1,1}\leq \alpha_{1,2}\leq \alpha_{1,3}$ if they are unordered. If the unrestricted MLEs are already ordered, they naturally coincide with the OR MLEs since the ladder are computed from the former. 

The OR MLEs produced average estimates which were closer to the true parameter value than the unrestricted MLEs in most cases, as can be seen from Table \ref{Table:Mean}. The table also demonstrates the unbiasdness of both estimates for larger sample sizes. For small sample sizes, MLEs may be non-existent or heavily overestimate the true model parameters. This issue also causes the inflated standard deviations. It has its origin in the small sample size that is present in some samples. However, they can be easily detected and considered suspicious (see Section 4 in \cite{Pesch2023}. for details). For those estimates computed largely free of this issue, the standard deviations of the OR MLEs were always smaller than those of their unrestricted counterparts.

\begin{table}[ht]
\centering
\caption{Mean, (standard deviations) and [proportion of existing estimates if not 100\%] of model parameters based on 10,000 Monte Carlo simulations.}
\label{Table:Mean}
\begin{tabular}{c|ccc} \hline
 $ r $ &  $\widehat{\alpha}_{1,1} $  &  $\widehat{\alpha}_{1,2} $  &  $ \widehat{\alpha}_{1,3} $  \\ \hline
5 & 2.97(2.33)[83.61\%] & 5.01(14.90)[81.94\%] & 11.87(104.53)[73.16\%] \\
10 & 2.27(1.31)[97.49\%] & 3.08(2.13)[96.56\%] & 4.79(29.31)[93.14\%]  \\
25 & 2.08(0.77) & 2.66(1.07)[99.98\%] & 3.07(1.47)[99.86\%]  \\ 
50 & 2.04(0.53) & 2.58(0.70) & 2.90(0.91)  \\ \hline
&  $ \widehat{\alpha}_{1,1}^* $ & $ \widehat{\alpha}_{1,2}^* $ & $\widehat{\alpha}_{1,3}^* $ \\ \hline
5 & 2.06(1.19) & 3.33(2.52) & 15.59(132.32)[44.62\%] \\
10 & 1.90(0.94) & 2.82(1.41) & 5.28(30.23) [87.28\%]\\
25 & 1.93 (0.61) & 2.59(0.76) & 3.39(1.37) [99.84\%] \\
50 &  1.96(0.45) & 2.52(0.52) & 3.08(0.79)  \\ \hline
\end{tabular}
\end{table}

\begin{figure}[ht]
\begin{centering}
\includegraphics[width=85mm]{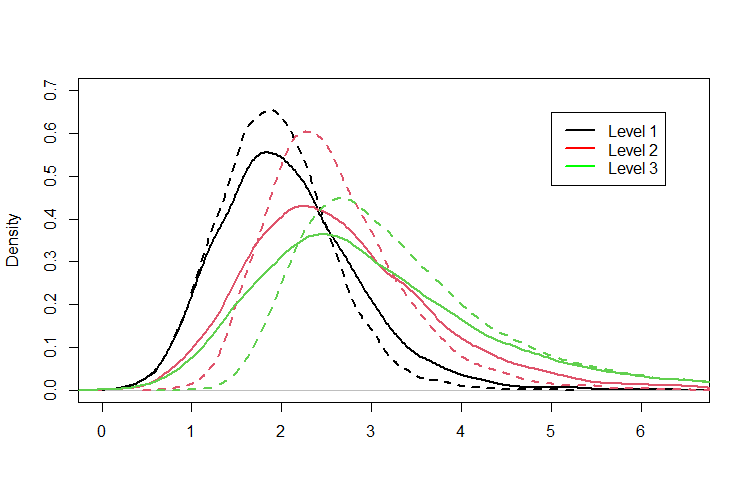}
\caption{Kernel density estimates based on the unrestricted MLEs (solid lines) and OR MLEs (dashed lines).}
\label{fig:Kernel}
\end{centering}
\end{figure}

Additionally, we obtained the kernel density estimates of all model parameters based on the same 10,000 MC simulations the mean and standard deviations were computed for, with $r=25$. Gaussian kernels with `rule of thumb' approach bandwidths were used for all computations. Figure \ref{fig:Kernel} displays the kernel densities and visualises the smaller standard deviations of the OR estimates. 
Summarizing, the presented results demonstrate that the OR MLEs are on average closer to the true value with smaller errors. 

The presented results clearly reveal the necessity for constrained methods of estimation. Table \ref{tab:FIRST} showcases how estimation can be improved by the use of OR MLEs for individual cases. Additonally, the OR MLEs were shown to be closer to the true parameter values on average with smaller deviations.

\subsection{Performance depending on the choice of parameter values}

In the previous section we demonstrated the usefulness of OR MLEs via a 3-out-of-4:F system with true parameters $\alpha_{1,1}=2,\,\alpha_{1,2}=2.5$ and $\alpha_{1,3}=2.75$ in regards to preserving the order restriction. Additional simulation studies were carried out to test how heavily the estimation results for an individual component depend on the true parameter values across the different levels. To this end, means and standard deviations were computed based on 10,000 MC simulations for different true parameter values and sample sizes $r\in\{5,10,25,50\}$. For the purpose of focusing on the effect that the choice of true parameter values has on the performance of these estimates, the remaining three components were simulated with the same sensitivity as component 1, i.e. $\alpha_{j,k}=\alpha_{1,k}$, $j\in\{2,3,4\},\,k\in\{1,2,3\}$. Meanwhile, the sensitivity differences across the three levels for component 1 are governed by a proportionality factor $p\in\{1,1.5,2\}$, i.e. $\alpha_{1,1}=1$ and $\alpha_{1,j}=p\,\alpha_{1,j-1},\,j\in\{2,3\}$. The results (see Table \ref{tab:Means}) clearly demonstrate that the OR MLEs appear to be robust against varying sensitivities across levels. Hence, the OR MLEs can be said to perform well independently of how quickly a component deteriorates homogeneously across levels.

\begin{table}[ht]
\centering
\caption{Mean, (standard deviation) and proportion of existing estimates of model parameters based on 10,000 Monte Carlo simulations.}
\label{tab:Means}
\begin{tabular}{c|cccc}
p=1 & $\alpha_{1,1}=1$ & $\alpha_{1,2}=1$ & $\alpha_{1,3}=1$ & prop. \\
r & $\widehat{\alpha}_{1,1}$ & $\widehat{\alpha}_{1,2}$ & $\widehat{\alpha}_{1,3}$ &  \\ \hline
5 & 1.07 0.64) & 1.49 (1.01) & 4.70 (53.84) & 38.14\% \\
10 & 0.88 (0.43) & 1.14 (0.55) & 1.65 (2.65) & 83.01\% \\
25 & 0.88 (0.28) & 1.04 (0.29) & 1.24 (0.42) & 99.80\% \\
50 & 0.91 (0.20) & 1.02 (0.20) & 1.14 (0.25) & 100\% \\ \hline
p=1.5 & $\alpha_{1,1}=1$ & $\alpha_{1,2}=1.5$ & $\alpha_{1,3}=2.25$ &  \\
r & $\widehat{\alpha}_{1,1}$ & $\widehat{\alpha}_{1,2}$ & $\widehat{\alpha}_{1,3}$ &  \\ \hline
5 & 1.26 (0.75) & 2.28 (1.60) & 10.40 (69.93) & 37.44\% \\
10 & 1.02 (0.54) & 1.73 (0.90) & 3.37 (5.08) & 83.45\% \\
25 & 0.99 (0.37) & 1.57 (0.53) & 2.54 (1.01) & 99.71\% \\
50 & 1.00 (0.27) & 1.54 (0.39) & 2.37 (0.65) & 100\% \\ \hline
p=2 & $\alpha_{1,1}=1$ & $\alpha_{1,2}=2$ & $\alpha_{1,3}=4$ &  \\
r & $\widehat{\alpha}_{1,1}$ & $\widehat{\alpha}_{1,2}$ & $\widehat{\alpha}_{1,3}$ &  \\ \hline
5 & 1.34 (0.85) & 3.14 (2.46) & 18.18 (127.66) & 38.32\% \\
10 & 1.08 (0.61) & 2.35 (1.32) & 5.95 (11.87) & 83.51\% \\
25 & 1.02 (0.40) & 2.11 (0.80) & 4.43 (1.90) & 99.81\% \\
50 & 1.02 (0.29) & 2.05 (0.57) & 4.19 (1.20) & 100\%
\end{tabular}
\end{table}

\begin{figure}[ht]
\begin{centering}
\includegraphics[width=85mm]{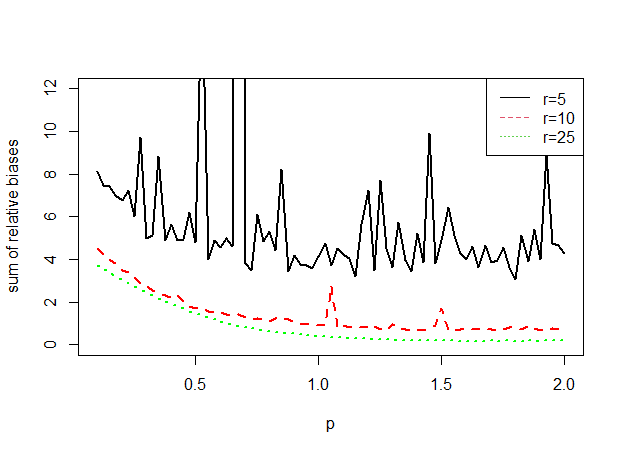}
\caption{Estimation accuracy based on level wise differences in true parameter values.}
\label{Fig:accuracy}
\end{centering}
\end{figure}

To further illustrate this point, we repeated these simulations for $p\in [0.1,2]$ with increments of $0.025$. Note values $p<1$ add the additional feature of decreasing risks to the analysis. The sum of the three relative biases across levels was used as a measure of estimation accuracy. Figure \ref{Fig:accuracy} confirms that performance does not depend on the choice of true parameters for $p\geq 1$. For $p<1$ however, it is obvious that the estimates perform poorly. This is not surprising, since by design they must follow an ascending order that is not present in the true values. Figure \ref{Fig:accuracy} also highlights, that large deviations from the true values become more unlikely with an increase in sample size.

\begin{figure}[ht]
\begin{centering}
\includegraphics[width=85mm]{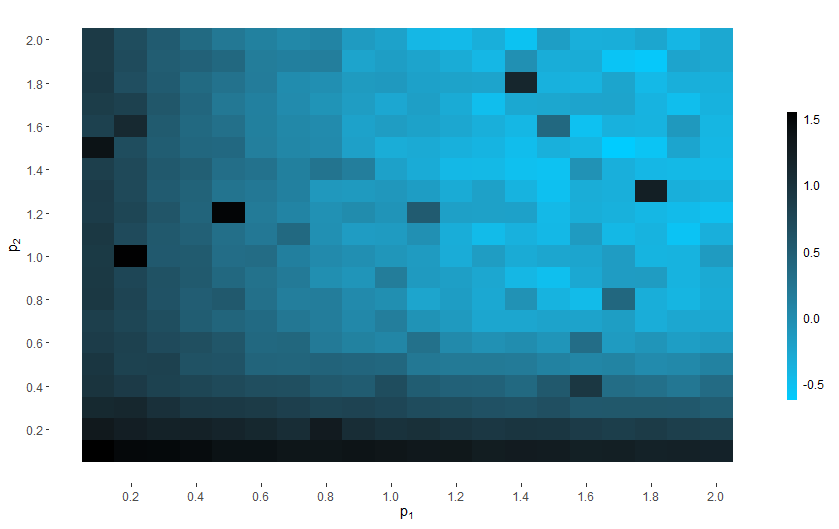}
\caption{Log scaled sum of relative biases for different proportionality factors $p_1,\,p_2 \in [0.1,2]$ between the three levels.}
\label{Fig:HEAT1}
\end{centering}
\end{figure}

\begin{figure}[ht]
\begin{centering}
\includegraphics[width=85mm]{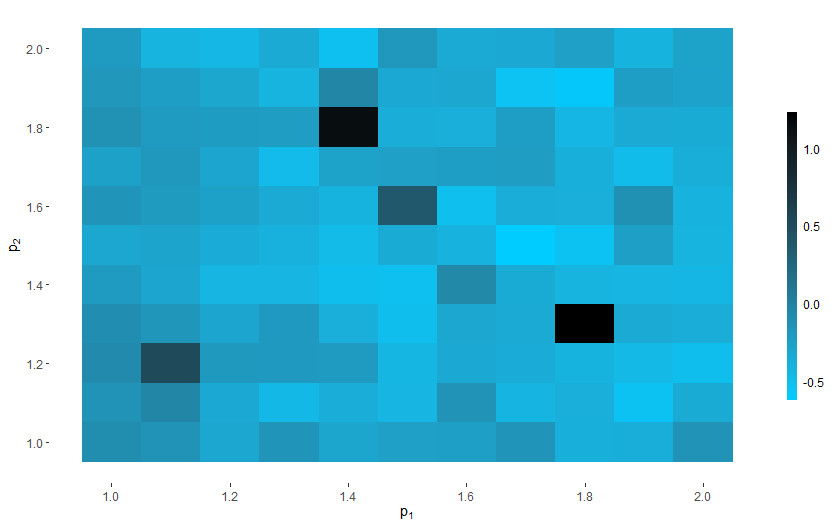}
\caption{Log scaled sum of relative biases for different proportionality factors $p_1,\,p_2 \in [1,2]$ between the three levels.}
\label{Fig:HEAT2}
\end{centering}
\end{figure}

We further modelled varying heterogeneous deterioration rates across levels. For component 1 only, the two proportionality factors $p_1$ and $p_2$ describe the deterioration of the component moving from level 1 to level 2 and from level 2 to level 3 respectively, i.e. $\alpha_{1,2}=p_1 \alpha_{1,1}$ and $\alpha_{1,3}=p_2 \alpha_{1,2}$, with $\alpha_{1,1}=1$. The color dimension of Figures \ref{Fig:HEAT1} and \ref{Fig:HEAT2} measure the log-scaled sum of relative biases for the estimation of all three model parameters based on $r=10$ observations. Figure \ref{Fig:HEAT1} illustrates once again, that the proposed OR MLEs behave poorly if failure risks decrease, i.e. $p_1,\,p_2 < 1$. This can be seen by the overall darker shades of blue as either $p_1$ or $p_2$ decrease. The higher the rate of decreasing risks, the worse the OR MLEs appear to perform. Figure \ref{Fig:HEAT2} depicts no clear color pattern and thereby not only confirms the robustness of the estimates against different homogeneous deterioration speeds as already seen in Figure \ref{Fig:accuracy} but also illustrates their robustness against varying heterogeneous deterioration rates between levels. The randomly scattered darker colored patches in both plots additionally highlight the chance of randomly inflated relative biases. This effect is based on the possibility of randomly observed `bad data' related to the small sample size as explained in Section \ref{Sec-4}. It is independent of $p_1$ and $p_2$ as can be seen from Figure \ref{Fig:HEAT2}. Similarly to the vanishing spikes in Figure \ref{Fig:accuracy}, the effect also vanishes for larger sample sizes, which we omit to graphically illustrate here.

Finally, we further investigate the relation of the sample size and the proportion of non-existing estimates. It is not surprising of course that this proportion increases as the number of observed systems goes up. The last column of Table \ref{tab:Means} additionally highlights that the proportion of obtainable estimates is independent of the increase in failure risk. However, the relation is mediated by other factors including the system size $n$ and level depth $s$ for example. If either of them increases, the proportion of existing estimates naturally decreases as less (or no) data is available for the calculation of the individual MLEs. We omit a more detailed investigation but focus on component comparability instead. It is reasonable to assume the proportion of existing estimates to decrease as component sensitivities deviate heavily between components. To verify this presumption, we used 10,000 MC simulations for a 3-out-of-4:F systems with $p=1.5$ as within-component proportionality factor for all components. To model varying between-component sensitivities, we used another proportionality factor $\widetilde{p}$ with $\alpha_{j,1}=\widetilde{p}\, \alpha_{j-1,1},\,j\in\{2,3,4\}$ and $\alpha_{1,1}=1$.

\begin{table}[ht]
\centering
\caption{Proportion of existing estimates}
\label{Tab:AscertEst}
\begin{tabular}{c|ccc}
$\widetilde{p}$ & $r=5$ & $r=10$ & $r=25$ \\ \hline
1 & 0.3820 & 0.8334 & 0.9981 \\
1.5 & 0.3572 & 0.7629 & 0.9838 \\
2 & 0.3068 & 0.6507 & 0.9259
\end{tabular}
\end{table}

\begin{figure}[ht]
\begin{centering}
\includegraphics[width=85mm]{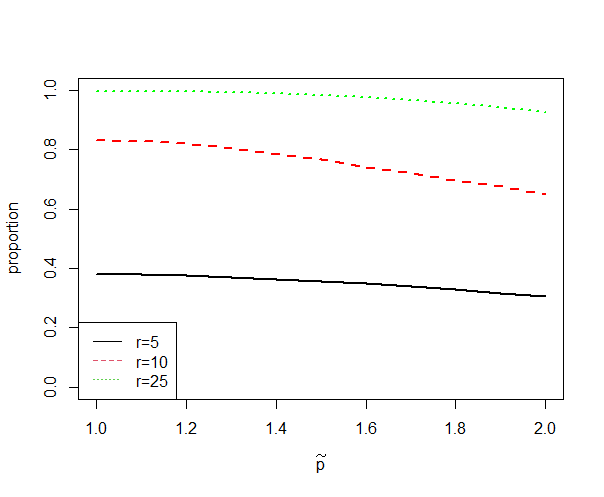}
\caption{Proportion of existing OR estimates for different between-component proportionality factors $\tilde{p}$.}
\label{Fig:AscertEst}
\end{centering}
\end{figure}

With these configurations and without loss of generality component 1 is hence modelled as the least and component 4 as the most sensitive on all respective levels. We computed the proportion of existing model estimates of the entire system. Table \ref{Tab:AscertEst} confirms the presumption. Independently of the sample size, the proportion of existing estimates decreases as the between component proportionality factor $\widetilde{p}$ increases, i.e. as components sensitivities become more different from another. To illustrate this effect further, we repeated the simulations for $\widetilde{p}\in[1,2]$ as can be seen in Figure \ref{Fig:AscertEst} which also supports the findings.

\subsection{Likelihood Ratio Test}
\label{Sec-6}

The usefulness of OR estimates has already been demonstrated extensively and is immediately apparent when considering successively increasing failure risks (see Example \ref{ex-ORvio}). We demonstrate another powerful application of OR MLEs. To this end, refer to the likelihood ratio test (LRT) introduced in \cite{Pesch2023}, which compares the model of SOS to the more general ESOS model. Under Assumption \ref{Ass-EQUALLOAD} the test favours the model of ESOS, if the null hypothesis
\[
H_0: F_{j,k} = \widetilde{F}_k,  j\in \{1,\dots,n\},\, k\in \{1,\dots,s\}
\]
is rejected, where $\widetilde{F}_k , k \in \{1,\dots,s\}$ are the underpinning model cdfs in the ordinary case (see \cite{Cramer1996}). In the following, further assume 
Assumption \ref{Ass-EQUALLOAD}, i. e.
\[
F_{j,k} = 1-(1-F_j^*)^{\alpha_{j,k}},
\]
where the baseline cdfs $F_j^*,\,j\in \{1,\dots,n\}$ are assumed known. Then, the restricted model features $s$ many model parameters; one for each level. The full model on the other hand considers $n\cdot s$ many parameters; one for each component on each level. The null hypothesis translates to
\[
H_0: \alpha_{j,k} = \alpha_k, j\in \{1,\dots,n\},\, k\in \{1,\dots,s\},
\]
where $\alpha_k$ are the model parameters in the ordinary SOS model.

For a system with $n$ components and $r$ samples $\big(X_{*l}^{(1)},C_{1,l},X_{*l}^{(2)},C_{2,l}\dots,X_{*l}^{(s)},C_{s,l}\big)_{1\leq l \leq r}$ of the first $s$ ESOS and failure sources, the LRT proceeds as follows:

\begin{enumerate}
\item 
Simulate the test statistic 
\[T = -2 \Big(\log \big(\sup_{\boldsymbol{\theta}\in \R_{+}^{s}}\, L(\boldsymbol{\alpha})\big) - \log \big(\sup_{\boldsymbol{\theta}\in \R_{+}^{ns}}\, L(\boldsymbol{\alpha})\big)\Big)
\]
under $H_0$ 100,000 times for varying values of $\boldsymbol{\alpha} \in \R_{+}^{s}$. Derive the exact quantiles of $T$ based on the simulation results.

\item Compute the $n \cdot s$ OR MLEs $\widehat{\alpha}_{j,k}^{*}$ as described in Theorem \ref{Theo-1} and substitute these into \eqref{eq-LLF}, i.e., compute $\sup_{\boldsymbol{\alpha} \in \R_{+}^{n \cdot s}} ~ L(\boldsymbol{\alpha})$.

\item Compute the $s$ OR MLEs $\widehat{{\alpha}}_{k},\,k\in \{1,\dots,s\}$ as described in Theorem 3.1 of \cite{Balakrishnan2008}. Based on their level, identify all parameters in $L(\boldsymbol{\alpha})$ accordingly. Then substitute these into $\eqref{eq-LLF}$, i.e., compute $\sup_{\boldsymbol{\alpha}\in \R_+^{s}}\, L(\boldsymbol{\theta})$.

\item Compute $T$ and compare the statistic to the exact ${(1-\alpha)\%}$ quantile from step 1 to arrive at a test decision.
\end{enumerate}

\begin{figure}[ht]
\begin{centering}
\includegraphics[width=85mm]{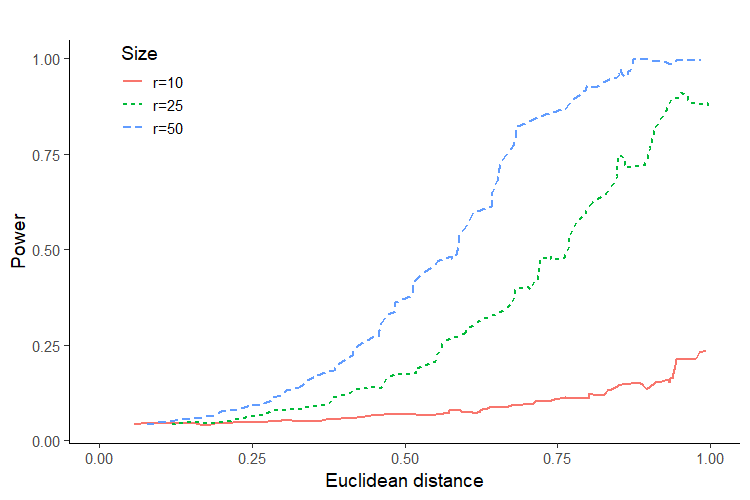}
\caption{Power for different sample sizes and different Euclidean distances to null based on varying true parameter values on level 1.}
\label{fig:Power1}
\end{centering}
\end{figure}

Two power studies for a 2-out-of-3:F system that was observed until failure illustrate the performance of the proposed LRT for different sample sizes. In the first study, the three model parameters of level 1 were randomly chosen between $0.1$ and $2$ a total of 100 times. For each combination, we ran 1,000 Monte Carlo simulations to obtain the empirical power of the above described LRT as the proportion of significant results. Each replication of the test was based on $r\in\{10,25,50\}$ many simulated ESOS. The level 2 parameters were all set to the maximum of the level 1 parameters. Differences between the components were hence only simulated to occur on level 1.
Figure \ref{fig:Power1} displays the running mean with window 5 of the power, based on the distance to the null hypothesis, which varies depending on the random choice of true parameter values.

\begin{figure}[H]
\begin{centering}
\includegraphics[width=85mm]{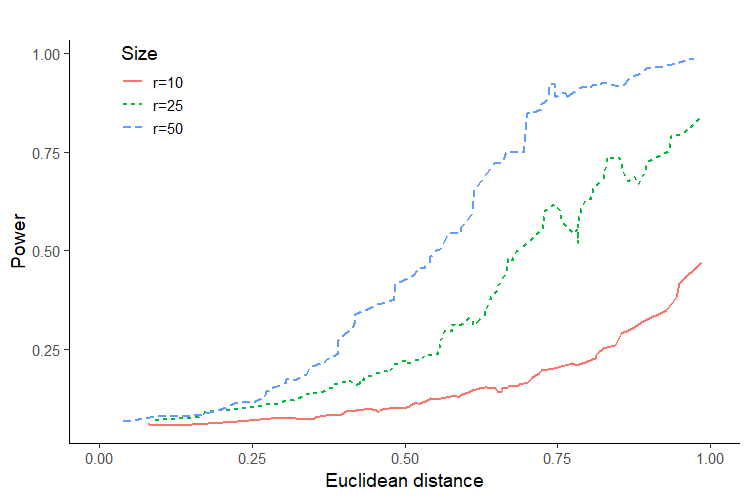}
\caption{Power for different sample sizes and different Euclidean distances to null based on varying true parameter values on level 2.}
\label{fig:Power2}
\end{centering}
\end{figure}

Similarly, in the second study, the three level 1 model parameters were all set to 0.5 and the level 2 parameters randomly chosen between 0.5 and 4. Differences between the components were hence only simulated to occur on level 2. Again, 100 random combinations were considered and power was calculated based on 1,000 simulations of the proposed LRT for different sample sizes.
Note how both of these studies are conservative in the sense that differences between components only occur on one level.
Figure \ref{fig:Power2} displays the power based on Euclidean distance to the null.

As expected the power converges to $5\%$ for distances approaching 0 regarding both levels. Both figures further demonstrate that an increase in sample size also increases the power. The same is true for increasing distances to the null. These results were to be expected and confirm the usefulness of the OR MLEs.

\section{Some parametric families of distributions}
\label{Sec-5}

Both, unrestricted and order restricted MLEs depend on the baseline distributions. If the baseline distributions are unknown, they also need to be estimated. In this case, let $F_j^*$ be a cdf of a parametric family of distributions ${\mathcal{P}_j=\{F^*_{\boldsymbol{\theta}_j}\mid \boldsymbol{\theta}_j \in \boldsymbol{\Theta}_j\}}\, {j \in\{1,\dots,n\}}$. Focus on one arbitrary component only, say component $\ell \in \{1,\dots,n\}$. Then, via the plug-in method the profile log-likelihood function $l^*(\boldsymbol{\theta}_\ell,\widehat{\boldsymbol{\alpha}}_{\ell})$ needs to be maximised with respect to the distribution parameter vector $\boldsymbol{\theta}_\ell$. 

Case 1: A scale family of distributions

For component $\ell$ consider a baseline cdf of the form $F_\ell^*(t)=1-e^{-\lambda g(t)},\,t\geq 0, \,\lambda>0,$ where $g$ is a known, increasing and differentiable function on $[0,\infty)$ satisfying $g(0)=0$ and $\lim_{t \rightarrow \infty} g(t) = \infty$. Then, the profile log-likelihood reads
\begin{align*}
&l^*(\lambda,\widehat{\boldsymbol{\alpha}}_\ell)=
\sum_{i=1}^r
\sum_{k=1}^s \big( \ln(g'(x_{k,i})) \big) \mathds{1}_{\{\ell\}}(c_{k,i}) -  \sum_{k=1}^s m_{\ell,k}+ \sum_{k=1}^s m_{\ell,k}\Bigg(\ln(m_{\ell,k})-\ln\bigg(\sum_{i=1}^r\eta_{k,i} (I_{\ell,k-1})_i\bigg)\Bigg),
\end{align*}

where $\eta_{k,i}=g(x_{k,i})-g(x_{k-1,i})$ and $g'(t)=\frac{\partial}{\partial t}g(t)$. The profile log-likelihood is independent of $\lambda$. Hence, no MLE for $\lambda$ exists but instead the classic identifiability problem occurs, encountered when estimating a product of parameters. Ergo, only $\widetilde{\alpha}_{\ell,k}=\lambda\alpha_{\ell,k}$ can be estimated. To this end, from Lemma \ref{Lemma-Alpha} and by the invariance property of MLEs,
\begin{align}
\label{eq-MLECase3}
\widehat{\alpha}_{c,k}=\frac{m_{c,k}}{\lambda \sum_{i=1}^r \eta_{k,i} (I_{c,k-1})_i}\Leftrightarrow  \,\widehat{\widetilde{\alpha}}_{c,k}= \frac{m_{c,k}}{\sum_{i=1}^r\eta_{k,i} (I_{c,k-1})_i}
\end{align}
are the associated MLEs of $\widetilde{\alpha}_{c,k}$.

For the choice $g(t)=t$, the exponential case is generated with MLEs
\[
\widehat{\widetilde{\alpha}}_{c,k}= \frac{m_{c,k}}{\sum_{i=1}^r (x_{k,i}-x_{k-1,i})(I_{c,k-1})_i}.
\]
\begin{Rem}
Other known distributions are also included in this family of distributions. For examples, the choices $g(t)=t^a,\, a>0$ and $g(t)=\ln(t),\, t\geq 1$ respectively, leads to Weibull and Pareto distributions, respectively.
\end{Rem}

Case 2: A location-scale family of distributions

\begin{Ass}
\label{Ass-GeneralFamiliy}
Consider baseline cdfs of the form
\[
F_\ell^*(t)= 1-e^{-\lambda(g(t)-\mu)},\hspace{2em} t\geq g^{-1}(\mu)
\]
where $\lambda > 0$, $\mu \in \R$ and $g:[a,b]\rightarrow \R$ with $-\infty \leq a < b \leq \infty$ and $g(a)=-\infty$ as well as $g(b)=\infty$ is a known, increasing and differentiable function. By convention, let $x_{0,i}=g^{-1}(\mu),\, 1\leq i \leq r$. 
\end{Ass}

Then, the profile log-likelihood reads
\begin{align*}
l^*(\lambda,\mu,&\widehat{\boldsymbol{\alpha}}_\ell)=
\sum_{i=1}^r
\sum_{k=1}^s \big( \ln(g'(x_{k,i})) \big) \mathds{1}_{\{\ell\}}(c_{k,i}) -  \sum_{k=1}^s m_{\ell,k}+ \sum_{k=1}^s m_{\ell,k}\Bigg(\ln(m_{\ell,k})-\ln\bigg(\sum_{i=1}^r\eta_{k,i} (I_{\ell,k-1})_i\bigg)\Bigg),
\end{align*}
which is independent of $\lambda$, so that the MLE of $\lambda$ does not exist. However, MLEs for $\widetilde{\alpha}_{\ell,k}=\lambda\alpha_{\ell,k}$ exists according to \eqref{eq-MLECase3}. Further, since $\mu\leq \min_{1\leq i \leq r} g(x_{1,i})$ and the likelihood function \eqref{eq-LLF} is increasing in $\mu$ for all fixed $\alpha_{c,k},\,1\leq k\leq s$, the MLE of $\mu$ is given by $\widehat{\mu}=\min_{1\leq i \leq r} g(x_{1,i})$. The results motivate the following theorem.

\begin{Theo}
\label{theo-a}
If a baseline cdf $F_j^*,\,j\in \{1,\dots,n\}$ satisfies Assumption \ref{Ass-GeneralFamiliy}, the MLEs of $\mu$ and $\widetilde{\alpha}_{j,k}=\lambda\alpha_{j,k}$
(under the order restriction $\widetilde{\alpha}_{j,1}\leq \dots \leq \widetilde{\alpha}_{j,s})$ are given by: 
\begin{align*}
\widehat{\mu}=\min_{1\leq i \leq r} g(x_{1,i})
\end{align*}
and
\begin{align*}
\widehat{\widetilde{\alpha}}_{j,s-k+1}^* = \bigg(\max_{l \leq k} \min_{t \geq k} \frac{\sum_{\nu=l}^t \widehat{\widetilde{\alpha}}_{j,s-\nu +1}^{-1} m_{j,s-\nu+1}}{\sum_{\nu=l}^t m_{j,s-\nu+1}}\bigg)^{-1}, 
\end{align*}
$1 \leq k \leq s$, where $\widehat{\widetilde{\alpha}}_{\ell,s}$ is given by representation $\eqref{eq-MLECase3}$.
\end{Theo}

\textit{Proof}: 

Consider the baseline distribution $F_j^*(t)= 1-e^{-\lambda(g(t)-\mu)}$. Then, the likelihood function (see \eqref{eq-LLF})
\begin{align*}
L(\mu,\boldsymbol{\widetilde{\alpha}}_\ell)&=
 K \,\prod_{k=1}^s\, \prod_{j\in \mathcal{N}} \alpha_{j,k}^{m_{j,k}}\,\, \prod_{i=1}^r \, \prod_{k=1}^s \, \prod_{j\in B_{\pi_{k-1,i}}} e^{-\widetilde{\alpha}_{j,k}\,\eta_{k,i}}\\
 &= 
 \widetilde{K} \, \prod_{i=1}^r \,\,\prod_{k=1}^s \, \prod_{j\in B_{\pi_{k-1,i}}} e^{-\widetilde{\alpha}_{j,k}\,\eta_{k,i}}\\
 &=\widetilde{\widetilde{K}}\,\, \prod_{i=1}^r \prod_{j\in \calN} e^{-\widetilde{\alpha}_{j,1} \big(g(x_{1,i})-\mu\big)},
\end{align*}
with $\widetilde{\widetilde{K}}>0$, is increasing in $\mu$ for all fixed $\boldsymbol{\widetilde{\alpha}}_\ell$. The proposed representation of $\widehat{\mu}$ then follows due to $\mu \leq \min_{1\leq i\leq r} g(x_{1,i})$.

Further, due to Lemma \ref{Lemma-Alpha} the unrestricted MLE of $\widetilde{\alpha}_{j,k}$ can readily be obtained via \eqref{eq-MLECase3} regardless of the shift parameter $\mu$. With Theorem \ref{Theo-1}, the OR MLEs of $\widetilde{\alpha}_{\ell,k}$ then follow directly.

\begin{flushright}
$\square$
\end{flushright}

\section{Conclusion}
We have addressed the goal of adequately modelling the assumption of increasing failure risks for surviving components in load sharing systems with heterogeneous components. To this end, we introduced a simple order restriction to the model of ESOS and proposed new estimates called OR MLEs. If the unrestricted MLEs adhere to the order restriction, the two types of estimates coincide. In situations where the order restriction is violated by the ordinary MLEs however, we demonstrated that the OR MLEs can serve as a valuable alternative as they meet the required restriction still. Since the violation of the order restriction becomes more likely with smaller sample sizes, the newly proposed estimates can prove particularly useful when data is scarce. It should also be mentioned however, that the OR MLEs may not be available at all for smaller sample sizes depending on the data. While it is obvious that a decrease in sample size also decreases the chance of existing estimates, establishing an exact relation between the proportion of existing estimates and the sample size appears complex since other factors, such as the level depths and the similarity of component parameters also affect this relation. While OR estimates work well for systems that have increasing risk of failure, they are also shown to be otherwise inappropriate. For this reason, we only recommend their use if system operators are sure that the physical properties of the system correspond to the assumption of increasing failure risks.

\section*{Acknowledgements}
This work was supported by the Australian Research Council through the Centre for Transforming Maintenance through Data Science  (grant number IC180100030), funded by the Australian Government.


\begin{thebibliography}{25}
\providecommand{\natexlab}[1]{#1}
\providecommand{\url}[1]{\texttt{#1}}
\expandafter\ifx\csname urlstyle\endcsname\relax
  \providecommand{\doi}[1]{doi: #1}\else
  \providecommand{\doi}{doi: \begingroup \urlstyle{rm}\Url}\fi

\bibitem[Kamps(1995)]{Kamps1995}
U.~Kamps.
\newblock {A concept of generalized order statistic}.
\newblock \emph{Journal of Statistical Planning and Inference}, 48\penalty0
  (1):\penalty0 1--23, 1995.
\newblock ISSN 03783758.
\newblock \doi{10.1016/0378-3758(94)00147-N}.

\bibitem[Zhao et~al.(2018)Zhao, Liu, and Liu]{Zhao2018}
X.~Zhao, B.~Liu, and Y.~Liu.
\newblock {Reliability Modeling and Analysis of Load-Sharing Systems with
  Continuously Degrading Components}.
\newblock \emph{IEEE Transactions on Reliability}, 67\penalty0 (3):\penalty0
  1096--1110, 2018.
\newblock ISSN 00189529.
\newblock \doi{10.1109/TR.2018.2846649}.

\bibitem[Sutar and Naik-Nimbalkar(2014)]{Sutar2014}
S.~S. Sutar and U.~V. Naik-Nimbalkar.
\newblock {Accelerated failure time models for load sharing systems}.
\newblock \emph{IEEE Transactions on Reliability}, 63\penalty0 (3):\penalty0
  706--714, 2014.
\newblock ISSN 00189529.
\newblock \doi{10.1109/TR.2014.2313793}.

\bibitem[Müller and Meyer(2022)]{Mueller2022}
C.~H. Müller and R.~Meyer.
\newblock Inference of intensity-based models for load-sharing systems with
  damage accumulation.
\newblock \emph{IEEE Transactions on Reliability}, 71\penalty0 (2):\penalty0
  539--554, 2022.
\newblock \doi{10.1109/TR.2022.3140483}.

\bibitem[Balakrishnan et~al.(2011)Balakrishnan, Beutner, and
  Kamps]{Balakrishnan2011}
N.~Balakrishnan, E.~Beutner, and U.~Kamps.
\newblock {Modeling parameters of a load-sharing system through link functions
  in sequential order statistics models and associated inference}.
\newblock \emph{IEEE Transactions on Reliability}, 60\penalty0 (3):\penalty0
  605--611, 2011.
\newblock ISSN 00189529.
\newblock \doi{10.1109/TR.2011.2161152}.

\bibitem[Bedbur et~al.(2019)Bedbur, Johnen, and Kamps]{Bedbur2019}
S.~Bedbur, M.~Johnen, and U.~Kamps.
\newblock Inference from multiple samples of weibull sequential order
  statistics.
\newblock \emph{Journal of Multivariate Analysis}, 169:\penalty0 381--399,
  2019.
\newblock ISSN 0047-259X.
\newblock \doi{https://doi.org/10.1016/j.jmva.2018.10.010}.
\newblock URL
  \url{https://www.sciencedirect.com/science/article/pii/S0047259X18300162}.

\bibitem[Mies and Bedbur(2019)]{Mies2019}
F.~Mies and S.~Bedbur.
\newblock Exact semiparametric inference and model selection for load-sharing
  systems.
\newblock \emph{IEEE Transactions on Reliability}, 69\penalty0 (3):\penalty0
  863--872, 2019.

\bibitem[Baratnia and Doostparast(2017)]{Baratnia2017}
M.~Baratnia and M.~Doostparast.
\newblock {Modeling lifetime of sequential r-out-of-n systems with independent
  and heterogeneous components}.
\newblock \emph{Communications in Statistics: Simulation and Computation},
  46\penalty0 (9):\penalty0 7365--7375, 2017.
\newblock ISSN 15324141.
\newblock \doi{10.1080/03610918.2016.1236956}.
\newblock URL \url{https://doi.org/10.1080/03610918.2016.1236956}.

\bibitem[Pesch et~al.(2023)Pesch, Polpo, Cripps, and Cramer]{Pesch2023}
T.~Pesch, A.~Polpo, E.~Cripps, and E.~Cramer.
\newblock {Reliability inference with extended sequential order statistics}.
\newblock \emph{Applied Stochastic Models in Business and Industry}, \penalty0
  (October 2022):\penalty0 1--16, 2023.
\newblock ISSN 15264025.
\newblock \doi{10.1002/asmb.2764}.

\bibitem[Cramer and Kamps(2001{\natexlab{a}})]{Cramer1999d}
E.~Cramer and U.~Kamps.
\newblock Sequential $k$-out-of-$n$ systems.
\newblock In N.~Balakrishnan and C.~R. Rao, editors, \emph{Handbook of
  Statistics: Advances in Reliability Vol.~20}, chapter~12, pages 301--372.
  Elsevier, Amsterdam, 2001{\natexlab{a}}.

\bibitem[Cramer and Kamps(2001{\natexlab{b}})]{Cramer2001a}
E.~Cramer and U.~Kamps.
\newblock {Estimation with sequential order statistics from exponential
  distributions}.
\newblock \emph{Annals of the Institute of Statistical Mathematics},
  53\penalty0 (2):\penalty0 307--324, 2001{\natexlab{b}}.
\newblock ISSN 00203157.
\newblock \doi{10.1023/A:1012470706224}.

\bibitem[Cramer and Kamps(2003)]{Cramer2003}
E.~Cramer and U.~Kamps.
\newblock {Marginal distributions of sequential and generalized order
  statistics}.
\newblock \emph{Metrika}, 58\penalty0 (3):\penalty0 293--310, 2003.
\newblock ISSN 00261335.
\newblock \doi{10.1007/s001840300268}.

\bibitem[Scheuer(1988)]{Scheuer1988}
E.~M. Scheuer.
\newblock {Reliability of an m-out-of-n System When Component Failure Induces
  Higher Failure Rates in Survivors}.
\newblock \emph{IEEE Transactions on Reliability}, 37\penalty0 (1):\penalty0
  73--74, 1988.
\newblock ISSN 15581721.
\newblock \doi{10.1109/24.3717}.

\bibitem[Cramer and Kamps(1996)]{Cramer1996}
E.~Cramer and U.~Kamps.
\newblock {Sequential order statistics and k-out-of-n systems with sequentially
  adjusted failure rates}.
\newblock \emph{Annals of the Institute of Statistical Mathematics},
  48\penalty0 (3):\penalty0 535--549, sep 1996.
\newblock ISSN 0020-3157.
\newblock \doi{10.1007/BF00050853}.
\newblock URL \url{http://link.springer.com/10.1007/BF00050853}.

\bibitem[Balakrishnan et~al.(2008)Balakrishnan, Beutner, and
  Kamps]{Balakrishnan2008}
N.~Balakrishnan, E.~Beutner, and U.~Kamps.
\newblock {Order restricted inference for sequential k-out-of-n systems}.
\newblock \emph{Journal of Multivariate Analysis}, 99\penalty0 (7):\penalty0
  1489--1502, 2008.
\newblock ISSN 0047259X.
\newblock \doi{10.1016/j.jmva.2008.04.014}.

\bibitem[Kim and Kvam(2004)]{Kim2004}
H.~Kim and P.~H. Kvam.
\newblock {Reliability estimation based on system data with an unknown load
  share rule}.
\newblock \emph{Lifetime Data Analysis}, 10\penalty0 (1):\penalty0 83--94,
  2004.
\newblock ISSN 13807870.
\newblock \doi{10.1023/B:LIDA.0000019257.74138.b6}.

\bibitem[Burkschat et~al.(2010)Burkschat, Kamps, and Kateri]{Burkschat2010}
M.~Burkschat, U.~Kamps, and M.~Kateri.
\newblock {Sequential order statistics with an order statistics prior}.
\newblock \emph{Journal of Multivariate Analysis}, 101\penalty0 (8):\penalty0
  1826--1836, 2010.
\newblock ISSN 0047259X.
\newblock \doi{10.1016/j.jmva.2010.03.017}.
\newblock URL \url{http://dx.doi.org/10.1016/j.jmva.2010.03.017}.

\bibitem[Schenk et~al.(2011)Schenk, Burkschat, Cramer, and Kamps]{Schenk2011}
N.~Schenk, M.~Burkschat, E.~Cramer, and U.~Kamps.
\newblock {Bayesian estimation and prediction with multiply Type-II censored
  samples of sequential order statistics from one- and two-parameter
  exponential distributions}.
\newblock \emph{Journal of Statistical Planning and Inference}, 141\penalty0
  (4):\penalty0 1575--1587, 2011.
\newblock ISSN 03783758.
\newblock \doi{10.1016/j.jspi.2010.11.009}.
\newblock URL \url{http://dx.doi.org/10.1016/j.jspi.2010.11.009}.

\bibitem[Shafay et~al.(2014)Shafay, Balakrishnan, and Sultan]{Shafay2014}
A.~R. Shafay, N.~Balakrishnan, and K.~S. Sultan.
\newblock {Two-sample Bayesian prediction for sequential order statistics from
  exponential distribution based on multiply Type-II censored samples}.
\newblock \emph{Journal of Statistical Computation and Simulation}, 84\penalty0
  (3):\penalty0 526--544, 2014.
\newblock ISSN 00949655.
\newblock \doi{10.1080/00949655.2012.718779}.

\bibitem[Ahmadi et~al.(2018)Ahmadi, Rezaei, and Yousefzadeh]{Ahmadi2018}
K.~Ahmadi, M.~Rezaei, and F.~Yousefzadeh.
\newblock {Progressively Type-II censored competing risks data for exponential
  distributions based on sequential order statistics}.
\newblock \emph{Communications in Statistics: Simulation and Computation},
  47\penalty0 (5):\penalty0 1276--1296, 2018.
\newblock ISSN 15324141.
\newblock \doi{10.1080/03610918.2017.1310232}.
\newblock URL \url{https://doi.org/10.1080/03610918.2017.1310232}.

\bibitem[Tsai et~al.(2021)Tsai, Xin, and Kao]{Tsai2021}
T.~R. Tsai, H.~Xin, and C.~H. Kao.
\newblock {Bayesian estimation based on sequential order statistics for
  heterogeneous baseline gompertz distributions}.
\newblock \emph{Mathematics}, 9\penalty0 (2):\penalty0 1--21, 2021.
\newblock ISSN 22277390.
\newblock \doi{10.3390/math9020145}.

\bibitem[Liu(1998)]{Liu1998}
H.~Liu.
\newblock {Reliability of a load-sharing k-out-of-n:G system: Non-iid
  components with arbitrary distributions}.
\newblock \emph{IEEE Transactions on Reliability}, 47\penalty0 (3 PART
  1):\penalty0 279--284, 1998.
\newblock ISSN 00189529.
\newblock \doi{10.1109/24.740502}.

\bibitem[Balakrishnan(2007)]{Balakrishnan2007c}
N.~Balakrishnan.
\newblock Permanents, order statistics, outliers, and robustness.
\newblock \emph{Rev. Mat. Complut.}, 20:\penalty0 7--107, 2007.

\bibitem[Navarro(2021)]{Navarro2021}
J.~Navarro.
\newblock \emph{Introduction to System Reliability Theory}.
\newblock Springer, Cham, 2021.
\newblock ISBN 9783030869526.

\bibitem[Barlow et~al.(1972)Barlow, Bartholomew, Bremner, and
  Brunk]{barlow1972}
R.~E. Barlow, D.~J. Bartholomew, J.~M. Bremner, and H.~D. Brunk.
\newblock \emph{Statistical Inference under Order Restrictions}.
\newblock John Wiley \& Sons, New York, 1972.

\end{thebibliography}
\end{document}